\documentclass[onecolumn,reprint,amsmath,amssymb,nofootinbib]{revtex4}
\usepackage{ulem}
\usepackage{amssymb}
\usepackage{amsmath}
\usepackage{graphicx}
\usepackage{dcolumn}
\usepackage[colorlinks,urlcolor=blue,citecolor=blue,linkcolor=blue]{hyperref}
\usepackage{color,units}
\usepackage[dvipsnames]{xcolor} 
\usepackage{lineno}
\usepackage{xspace}
\usepackage{longtable} 
\usepackage{float} 
\usepackage{multirow}
\usepackage{amsfonts,wasysym,epsfig,verbatim,subfigure,bm,mathrsfs,lipsum}

\begin{document}
\newcommand{\IUCAA}{Inter-University Centre for Astronomy and
Astrophysics, Post Bag 4, Ganeshkhind, Pune 411 007, India}

\newcommand{\MPI}{Max-Planck-Institut f{\"u}r Gravitationsphysik (Albert-Einstein-Institut), D-30167 Hannover, Germany}

\newcommand{\LBNZ}{Leibniz Universit{\"a}t Hannover, D-30167 Hannover, Germany}

\title{Horizon fluxes of binary black holes in eccentric orbits}

\author{Sayak Datta}\email{sayak.datta@aei.mpg.de}
\affiliation{\MPI}\affiliation{\LBNZ}
\date{\today}
\newcommand{\sayak}[1]{{\color{red}\bf  SD:} {\color{red} #1}}
\newcommand{\jp}[1]{{\color{orange} #1}}

\begin{abstract}
We compute the rate of change of mass and angular momentum of a black hole, namely tidal heating, in an eccentric orbit. The change is caused due to the tidal field of the orbiting companion. We compute the result for both the spinning and non-spinning black holes in the leading order of the mean motion, namely $\xi$. We demonstrate that the rates get enhanced significantly for nonzero eccentricity. Since eccentricity in a binary evolves with time we also express the results in terms of an initial eccentricity and azimuthal frequency $\xi_{\phi}$. In the process, we developed a prescription that can be used to compute all physical quantities in a series expansion of initial eccentricity, $e_0$. These results are computed taking account of the spin of the binary components. The prescription can be used to compute very high-order corrections of initial eccentricity. We use it to find the contribution to eccentricity up to $\mathcal{O}(e_0^5)$ in the spinning binary. We also provide an approximate expression for $\mathcal{O}(e_0^n)$, where $n$ is any odd number. With this, we compute approximate expression for $\mathcal{O}(e_0^7)$ and $\mathcal{O}(e_0^9)$ for non-spinning binary. Using the computed expression of eccentricity,  we derived the rate of change of mass and angular momentum of a black hole, both rotating and non-rotating, in terms of initial eccentricity and azimuthal frequency up to $\mathcal{O}(e_0^6)$. We also compute leading order dephasing in both cases analytically up to $\mathcal{O}(e_0^6)$ and study its impact.
\end{abstract}

\maketitle

\section{Introduction}
\label{sec:introduction}

In recent times, the detection of gravitational waves (GWs) from the coalescence of compact binaries by LIGO~\cite{TheLIGOScientific:2014jea} and Virgo~\cite{TheVirgo:2014hva} has opened up a new era of astronomy~\cite{LIGOScientific:2018mvr, LIGOScientific:2020ibl}. As a result, these observations have motivated General Relativity (GR) tests ~\cite{LIGOScientific:2019fpa, Abbott:2018lct}. The components of the compact binaries observed by LIGO and Virgo are mainly inferred to be either black holes (BHs) or neutron stars (NSs), which is primarily based on the measurements of component masses, population models, and tidal deformability of NSs~\cite{Cardoso:2017cfl}. The merger of two NSs was also observed in the event GW170817~\cite{gw170817}, and possibly also GW190425~\cite{Abbott:2020uma}. More recently, detections of GW200105 and GW200115~\cite{bhns_LIGOScientific:2021qlt} were made where it is believed that it is BH-NS binary. However, it remains to be conclusively proven that the heavier objects observed are indeed BHs of GR.

Currently, the existing detectors are continuously being upgraded. Alongside, there are proposals for several next-generation ground-based detectors such as the Einstein telescope \cite{maggiore2020science} and cosmic explorer \cite{Reitze:2019iox}. These detectors will be significantly more sensitive detectors. As a result, it will be possible to measure very small features in the signals. Similarly, space-based detectors such as Laser Interferometer Space Antenna (LISA) \cite{LISA:2017pwj} are being built. LISA will observe binaries comprising supermassive bodies. These sources will be either very loud or will last very long in the detector for the detector to measure very small features in the signal. Therefore modeling the signals as accurately as possible has become a necessity.

Due to the causal structure, BHs in GR are perfect absorbers that behave as dissipative systems~\cite{MembraneParadigm, Damour_viscous, Poisson:2009di, Cardoso:2012zn}.  A striking feature of a BH is its horizon, which is a null surface that does not allow energy to escape outward. In the presence of a companion, the tidal field of the companion affects the BH. These tidal effects change BH's mass, angular momentum, and horizon area. This phenomenon is called {\it tidal heating} (TH)~\cite{Hartle:1973zz,Hughes:2001jr,PoissonWill}. If the BH is nonspinning, then energy and angular momentum flow into the BH. However, if the BHs are spinning then they can transfer their rotational energy from the ergoregion out into the orbit due to tidal interactions with their binary companion. Energy exchange via TH backreacts on the binary's evolution, resulting in a shift in the phase of the GWs emitted by the system. TH of objects such as NSs or horizonless ECOs is comparatively much less due to their lack of a horizon. So, a careful measurement of this phase shift can be used in principle to distinguish BHs from horizonless compact objects~\cite{Maselli:2017cmm, Datta:2019euh, Datta:2019epe, Datta:2020rvo, Agullo:2020hxe, Chakraborty:2021gdf, Sherf:2021ppp, Datta:2021row, Sago:2021iku, Maggio:2021uge, Sago:2022bbj}.

 A compact binary coalescence (CBC) consists primarily of three phases - inspiral, merger, and ringdown. The inspiral can be modeled using post-Newtonian (PN) formalism, whereas numerical relativity (NR) simulations are needed to model the merger regime~\cite{Pretorius:2007nq}. To study the ringdown part of the dynamics, one needs BH perturbation theory techniques~\cite{Sasaki:2003xr} or NR. Tidal heating is relevant in the inspiral and mostly in the late inspiral when the components are closer together making their tidal interactions stronger. In the PN regime, TH can be incorporated into the gravitational waveform by adding the phase and amplitude shift due to this effect into a PN approximant in the time or frequency domain.

In several studies, TH of black holes has been studied analytically or numerically. It has been demonstrated that in the extreme mass ratio inspirals the impact of TH will be significant \cite{Datta:2019epe, Datta:2019euh}. It has also been pointed out that TH can be used to distinguish between black holes and neutron stars. As a result, this can address the mass gap problem \cite{Datta:2020gem}. The observability in the third-generation detectors has been also studied \cite{Mukherjee:2022wws}. Since TH depends on the near horizon properties of a black hole, it is expected that any modification to the near horizon physics can lead to significant modification to TH. Therefore, it can be used to test the near-horizon properties of the black holes which may shed light on the quantum nature of gravity.

However, while computing the TH of classical black holes or studying its impact on the observation only the circular orbits have been considered. Currently, we lack a proper computation as well as an understanding of the TH of black holes in a generic orbit. In the current work, we would like to initiate bridging this gap by analytically computing the leading order TH of BHs in an eccentric orbit. We also compute the leading order dephasing due to eccentricity effects on TH. We will leave the computation of TH in a generic orbit for the future.

Throughout the article, we will use geometric units, assuming $G=c=1$ except when it will be required to demonstrate the post-Newtonian (PN) order. Latin indices represent spatial indices running from $1$ to $3$. All masses in the plots are in units of solar mass $M_{\odot}$.

\section{Tidal moments for a two-body system}
\label{sec:Tidal moments}

The post-Newtonian environment upto $1.5$PN order can be described by potentials $U_{ext}$, $U^a_{ext}$ and $\Psi_{ext}$ as discussed in Ref. \cite{Taylor:2008xy}. Once these potentials are known the motion of a BH in the barycentric frame $(t,x^a)$ can be determined \cite{Taylor:2008xy}. Similarly, the barycentric tidal moments, $\mathcal{E}_{ab}$ and $\mathcal{B}_{ab}$ can be computed from these potentials. TH is directly connected to the tidal moments ``perceived" by the BHs, namely $\Bar{\mathcal{E}}_{ab}$ and $\Bar{\mathcal{B}}_{ab}$. The corresponding transformation equations can be found in Ref. \cite{Taylor:2008xy}. Note, this prescription is valid for any post-Newtonian environment.

In the current work, we are interested in the post-Newtonian environment of a binary system. Therefore, the external environment is sourced by a single external body. Assuming the external source to be a post-Newtonian monopole of mass $m_2$ at position ${\bf z}_2$, the external potentials and their derivatives can be expressed in terms of $m_2$, the velocity of the body $v^a_2$, separation $r$ and the direction vector ${\bf n}$. These expressions can be further simplified if represented with respect to separation vector and relative velocity and by identifying the system's barycenter with the origin of the coordinate system.

For a generic post-Newtonian orbit fixed in a plane, we can identify the plane to be the $x-y$ plane. We then use the polar coordinates $r$ and $\phi$ to describe the orbital motions. In this system ${\bf r} = (r\cos\phi, r\sin\phi,0)$, and all vectors can be resolved in the basis ${\bf n} = (\cos\phi, \sin\phi,0)$, ${\bf \Phi} = (-\sin\phi, \cos\phi, 0)$, and ${\bf l} = (0, 0, 1)$ is the vector normal to the plane. In this system the tidal field can be written as \cite{Taylor:2008xy},

\begin{equation}
\label{eq:tidal moments}
    \begin{split}
        \mathcal{E}_{ab} =&  -\frac{3m_2}{r^3}{\bf n}_{\langle ab\rangle} -\frac{3m_2}{r^3} \Big(\left[-\frac{3m_1^2}{2m^2}\Dot{r}^2 + 2(r\Dot{\phi})^2 -\frac{5m_1 +6m_2}{2r}\right]{\bf n}_{\langle ab\rangle}\\
&-\frac{(2m_1 + m_2)m_2}{m^2}\Dot{r}(r\Dot{\phi}) {\bf n}_{\langle a} {\bf \Phi}_{b\rangle} + (r\Dot{\phi})^2 {\bf \Phi}_{\langle ab\rangle}\Big) + \mathcal{O}(c^{-4})\\
\mathcal{B}_{ab} =&  -\frac{6m_2}{r^3}(r\Dot{\phi}){\bf l}_{(a} {\bf n}_{b)} +\mathcal{O}(c^{-2}),
    \end{split}
\end{equation}

where $r$ is the interbody distance, $\phi$ is the angular position of the relative orbit in the plane. We use the notation ${\bf n}_{\langle ab\rangle} = {\bf n}_a {\bf n}_b - \frac{1}{3} \delta_{ab}$ and ${\bf \Phi}_{\langle ab\rangle} = {\bf \Phi}_a {\bf \Phi}_b - \frac{1}{3} \delta_{ab}$.

\section{Application to eccentric orbit}
\label{sec:eccentric intro}

To address the eccentric orbits we will here discuss the quasi-Keplerian formalism first introduced in Refs. \cite{Damour:1988mr, Schafer:1993abc}. It provides a solution to the conservative parts of the post-Newtonian (PN) equations of motion. For the current purpose, the orbital variables $(r,\phi)$ and their derivatives as a function of a parametric angle, namely the eccentric anomaly $u$ are required to be known. It can be achieved with quasi-Keplerian formalism. In this section, we will discuss the relevant physics of a binary in an eccentric orbit. This will set the premise for the computation of the rate of change of mass and angular momentum of a black hole as well as the evolution of the eccentricity of an orbit.

There are three angle variables involved in this approach. The angles $u,\,l$, and $v$ are the eccentric, mean, and true anomalies. The geometrical interpretation of these angles has been discussed in Ref. \cite{Moore:2016qxz}. The semimajor axis of the ellipse is $a_r=(M/n^2)^{1/3}$, with $M=m_1+m_2$. The mean motion is defined as, $n\equiv 2\pi/P$ with $P$ being the radial orbital (periastron to periastron) period. The current work focuses on deriving the horizon fluxes in leading order in dimensionless radial angular orbit frequency $\xi = M n$. We will use this dimensionless variable extensively in the current work since it helps in getting PN expansions.

In the presence of the post-Newtonian terms the orbits are not exact Keplerian ellipses. Their parametric equations for $r,\, \phi, \, \Dot{r}$, and $\Dot{\phi}$ still take a similar form. However, these relations are much more complicated. These resulting solutions are referred to as quasi-Keplerian. The explicit analytic expressions can be derived from the conservative contributions to the equations of motion. For this, the radiation-reaction contributions to the equations of motion are ignored.

Unlike in the Newtonian case, now three eccentricities $e_t$, $e_r$, and $e_{\phi}$ are required to describe the orbit instead of one. These eccentricities are related to each other and to the orbital energy and angular momentum \cite{Damour:2004bz, Memmesheimer:2004cv}. The quasi-Keplerian equations also show the well-known periastron precession. We will not show the expressions here for brevity. The expressions can be found in Ref. \cite{Moore:2016qxz}.

To find the required PN expansions appropriate set of constants of the motion needs to be chosen. There are several possible choices for the principal constants of motion \cite{Damour:2004bz, Memmesheimer:2004cv}. When radiation reaction is considered the evolution of these constants leads to the inspiral motion and the resulting complete GW waveforms. With the choice of a set of constants $r$, $\Dot{r}$, $\phi$, and $\Dot{\phi}$ is expressed in terms of $n$, $e_t$, $c_l$ $c_{\lambda}$ and the eccentric anomaly $u$ \cite{Memmesheimer:2004cv, Damour:2004bz}. For the current purpose, only $1.5$PN results in a harmonic gauge will suffice. For this reason, we demonstrate the $1.5$PN harmonic gauge expressions below,

\begin{widetext}
    \begin{eqnarray}
    \label{eq:orbit expression}
    r =& \frac{M}{\xi ^{2/3}} \left(1-e_t \cos u\right) \left(1+\frac{\xi ^{2/3}}{6 \left(1-e_t
   \cos u\right)} \left(e_t \left(7\eta-6\right) \cos u+2\eta-18\right)\right) \nonumber \\
    \Dot{r} =& \frac{e_t \xi^{1/3}\sin u}{1-e_t \cos u} \left(1+\frac{1}{6} \xi ^{2/3} \left(6-7\eta\right)\right) \\
    \Dot{\phi} =& \frac{\sqrt{1-e_t^2} \xi}{M \left(1-e_t \cos u\right){}^2}  \left(1+\frac{\xi ^{2/3}}{\left(1-e_t^2\right) \left(1-e_t \cos
   (u)\right)} \left(e_t \left(1-\eta\right) \cos u- e_t^2 \left(4-\eta\right) +3\right)\right).\nonumber\\
   \Dot{u} =& \frac{\xi}{M(1-e_t \cos (u))} + \mathcal{O}(\xi^{7/3}), \nonumber
\end{eqnarray}
\end{widetext}

where $\eta=m_1m_2/M^2$. The last equation has been derived after taking a time derivative of $l = n(t-t_0) +c_l = u-e_t \sin(u) +\mathcal{O}(\xi^{4/3})$. In the quasi-Keplerian formalism, PN expansions of elliptical orbit quantities are done in terms of the radial orbit angular frequency $\omega_r \equiv n \equiv \xi/M$. The limitation of this frequency is that correspondence with the circular orbit limit becomes tricky. In the circular orbit case it is more natural to expand in terms of the azimuthal $\omega_{\phi} \equiv \xi_{\phi}/M$. This frequency corresponds to the time taken by a body to return to the same azimuthal angle. Indeed these two frequencies are connected to each other \cite{Moore:2016qxz}. In the current work, we will use the relationship between the orbit-averaged azimuthal frequency $\omega_{\phi} \equiv \langle \Dot{\phi}\rangle$ and $\omega_r$. Using orbital averaging the dependence on eccentric anomalies can be averaged out. As a result, it is possible to express $\xi$ in terms $e_t$ and $\xi_{\phi}$ as follows \cite{Moore:2016qxz},

\begin{equation}
\label{eq:xi to xiphi}
        \frac{\xi}{\xi_{\phi}} = 1 - \frac{3\xi_{\phi}^{2/3}}{1-e_t^2} -\left[18 -28\eta+ \left(51 -26\eta \right)e_t^2 \right]\frac{\xi_{\phi}^{4/3}}{4(1-e_t^2)^2}
\end{equation}

where $\xi_{\phi} = \pi Mf$. $f$ is the GW frequency and $M=(m_1 + m_2)$.

In the $e_t \rightarrow 0$ limit the frequency variables $\xi$ and $\xi_{\phi}$ do not agree. This is due to the fact that the periastron advance angle is independent of eccentricity in the $e_t \rightarrow 0$ limit. Since $\xi_{\phi}$ the circular orbit frequency in the $e_t \rightarrow 0$ limit, while comparing the PN quantities in the $e_t \rightarrow 0$ limit, it is necessary to express them in terms of $\xi_{\phi}$ (not $\xi$). This is a crucial point and the reason behind computing all the quantities in the end in terms of $\xi_{\phi}$.

\section{Leading order tidal heating}
\label{sec:Leading order TH}

In this section, we will use the results from the previous sections and find the leading order rate of change of mass and angular momentum of a black hole immersed in a tidal field. To compute the rates it is required to go to the black holes frame. Quantities in the BH frame are represented with an overbar. The rates directly depend on the tidal fields defined on the BH frame. The definition and impact of such transformations can be found in \cite{Taylor:2008xy}. We will only show the required expressions here.

\begin{eqnarray}
\label{eq: tidal moments in changed coordinate}
    \Bar{\mathcal{E}}_{ab}(\Bar{t}) =& \mathcal{N}_a^c(t) \mathcal{N}_b^d(t)\mathcal{E}_{ab}(t)\\
    \Bar{\mathcal{B}}_{ab}(\Bar{t}) =& \mathcal{N}_a^c(t) \mathcal{N}_b^d(t)\mathcal{B}_{ab}(t),
\end{eqnarray}

where the transformation is defined by, 

\begin{eqnarray}
\label{eq:co ordinate change}
    \Bar{t} =& t - A(t)\\
    \mathcal{N}_{ab}(t) =& \delta_{ab} -\epsilon_{abc}R^c(t),
\end{eqnarray}
where $\epsilon_{abc}$ is the permutation symbol of ordinary vector calculus with $\epsilon_{123}=1$. The governing equation of $A(t)$ and $R^c(t)$ is \cite{Taylor:2008xy}

\begin{eqnarray}
\label{eq: coordinate changed variable}
    \Dot{A} =&   \frac{m_2^2}{2m^2}[\Dot{r}^2 + (r\Dot{\phi})^2] + \frac{m_2}{r} +\mathcal{O}(c^{-2})\\
\Dot{R}^a =& -\Omega l^a +\mathcal{O}(c^{-2})\\
\Omega =& \frac{m_2}{2r^2}\frac{4m_1 +3m_2}{m} (r\Dot{\phi}).
\end{eqnarray}

Under this transformation angular coordinate $\phi \rightarrow \Bar{\phi}$. These two angles are connected to each other via $R^a$ and as a result, the rate of change of the angles depends on $\Dot{R}^a$. We find,

\begin{equation}
\label{eq: rotation in eccentric case}
    \begin{split}
        l^a\Bar{\phi} =& l^a\phi + R^a\\
        \frac{d\Bar{\phi}}{d\Bar{t}} =& (\frac{d\phi}{dt} -\Omega)\frac{dt}{d\Bar{t}}.
    \end{split}
\end{equation}

\subsection{Non-rotating black hole}

Once the tidal fields are computed in the BH frame the fluxes can be directly computed from it. We will assume that the body labeled as one in the binary is a BH. Then for a non-rotating BH the relation was found to be \cite{Taylor:2008xy},

\begin{equation}
\label{eq:spinless mass rate}
    \Dot{m}_1 = \frac{16 (Gm_1)^6}{c^{15}45} \left(\Dot{\Bar{\mathcal{E}}}_{ab} \Dot{\Bar{\mathcal{E}}}^{ab}   +  \frac{1}{c^2}\Dot{\Bar{\mathcal{B}}}_{ab} \Dot{\Bar{\mathcal{B}}}^{ab} +\mathcal{O}(c^{-4})\right).
\end{equation}

The above expression can be simplified using the results in Sec. \ref{sec:Tidal moments} and substituting the orbital variables with Eq. (\ref{eq:orbit expression}). We substitute Eq. (\ref{eq:orbit expression}) in Eq. (\ref{eq:tidal moments}). Then using Eq. (\ref{eq:co ordinate change}), Eq. (\ref{eq: coordinate changed variable}) and Eq. (\ref{eq: rotation in eccentric case}) tidal moments in Eq. (\ref{eq: tidal moments in changed coordinate}) is computed. These results are substituted in Eq. (\ref{eq:spinless mass rate}) to find the rate of change of mass. The resulting expression depends on the eccentric anomaly $u$. As a result, the expression is not fixed over an orbital period, it changes along with the change in $u$. Conventionally this is addressed by taking an orbital average. To compute the average over any quantity, say $X(u)$,  we average over orbital time period $(T)$. Then using the relation between time and $u$ this averaging can be translated over to an average over $u$ as follows \cite{Peters:1963ux, Moore:2016qxz}:

\begin{equation}
\label{eq:avearaging definition}
    \langle X\rangle \equiv \int_0^T \frac{dt}{T}X=\int^{2 \pi}_0 \frac{d u}{2 \pi} (1-e_t \cos{u}) X(u).
\end{equation}

Using Eq. (\ref{eq:spinless mass rate}) and Eq. (\ref{eq:avearaging definition}) we find the rate of change of mass of a nonspinning BH to be,
\begin{equation}
  \langle\Dot{m}_1\rangle =  \mathcal{M}    
 \frac{32 m_1^6 m_2^2 \xi ^6}{5  M^8}
\end{equation}

\begin{equation}
    \mathcal{M} = \frac{\left(25 e_t^8+740 e_t^6+2040 e_t^4+992 e_t^2+64\right) }{64 \left(1-e_t^2\right)^{15/2}} 
\end{equation}

This is the leading order horizon energy flux of a non-rotating black hole in an eccentric orbit. Note, the enhancement factor $\mathcal{M} \rightarrow 1$ in the limit $e_t\rightarrow 0$. Then in the leading post-Newtonian order, i.e. $\xi =\xi_{\phi}$ this exactly matches with the results derived for circular orbits in Ref. \cite{Alvi:2001mx}. A similar result in extreme mass ratio inspiral (EMRI) limit was computed in \cite{Forseth_thesis}. However, these results bank only on black hole perturbation theory where the source is a moving point particle. On the contrary in the current approach, only the validity of the PN approach is assumed. Therefore, any tidal field respecting PN conditions can be used to compute results. As a result, our result is usable for any mass ratios. Assuming $m_1\geq m_2$ and $m_2=qm_1$, $\langle\Dot{m}_1\rangle \propto q^2/(1+q)^8$. In EMRI limit $(q\ll1)$ this becomes $\langle\Dot{m}_1\rangle \propto  q^2$. Since $\xi$ and $e_t$ both depend on $\xi_{\phi}$ the final result in terms of $\xi_{\phi}$ will require further computations. This aspect will be addressed in later sections.

\subsection{Rotating black hole}

In this section, we compute the leading order rate of change of mass and angular momentum of a spinning black hole in an eccentric orbit. The rate of change of angular momentum of a rotating black hole in the presence of a general tidal field was computed in \cite{Poisson:2004cw, Taylor:2008xy}.

\begin{equation}
\label{eq:mass rate in spinng case}
    \begin{split}
        \Dot{J}_1 =& -\frac{2(Gm_1)^5}{45c^{10}} \chi \Big(8(1+3\chi^2)(E_1 +c^{-2}B_1) -3(4+17\chi^2)(E_2+c^{-2}B_2) +15\chi^2 (E_3 + c^{-2}B_3) +\mathcal{O}(c^{-4})\Big) \\
        \Dot{m}_1 c^2=& \frac{d\Bar{\phi}}{d\Bar{t}}\Dot{J}_1,
    \end{split}
\end{equation}
where, $E_1 = \Bar{\mathcal{E}}_{ab} \Bar{\mathcal{E}}^{ab},\,\,E_2 = \Bar{\mathcal{E}}_{ab}s^b \Bar{\mathcal{E}}^{a}_cs^c,\,\, E_3 = (\Bar{\mathcal{E}}_{ab} s^a s^b)^2,\,\, B_1 = \Bar{\mathcal{B}}_{ab} \Bar{\mathcal{B}}^{ab},\,\,B_2 = \Bar{\mathcal{B}}_{ab}s^b \Bar{\mathcal{B}}^{a}_cs^c,\,\, B_3 = (\Bar{\mathcal{B}}_{ab} s^a s^b)^2$ and $\chi$ is dimensionless spin parameter. The unit vector $s^a$ is aligned with the direction of the hole's spin angular momentum vector, therefore ${\bf J}_1 = J_1 {\bf s}$. We will assume that the spin is either aligned or antialigned with respect to the orbital angular momentum. Therefore $s^a = \epsilon l^a$, where $\epsilon =\pm 1$ depending on the orientation. Check Ref. \cite{Saketh:2022xjb} for another approach. Evaluating this in a similar manner to the previous section, we substitute Eq. (\ref{eq:orbit expression}) in Eq. (\ref{eq:tidal moments}). Then using Eq. (\ref{eq:co ordinate change}), Eq. (\ref{eq: coordinate changed variable}) and Eq. (\ref{eq: rotation in eccentric case}) tidal moments in Eq. (\ref{eq: tidal moments in changed coordinate}) is computed. These results are substituted in Eq. (\ref{eq:mass rate in spinng case}) to find the rate of change of mass.

the leading order rate of change of angular momentum can be found to be,

\begin{equation}
   \begin{split}
       \Dot{J}_1 =& -\frac{8 m_1^5 m_2^2 \xi ^4 \chi  \left(3 \chi ^2+1\right)}{5 M^6 (e_t \cos (u)-1)^6} \\
   \Dot{m}_1 =& -\frac{8 \epsilon\sqrt{1-e_t^2} m_1^5 m_2^2 \xi ^5 \chi  \left(3 \chi ^2+1\right)}{5 M^7 (e_t \cos (u)-1)^8}
   \end{split}
\end{equation}

Similar to the non-rotating black hole fluxes of a rotating black hole in an eccentric orbit are not a fixed quantity. Since it depends on the eccentric anomaly $u$, it changes across the orbit. Hence, with the instantaneous results at hand, we can take the orbital average of the rates as discussed in Eq. (\ref{eq:avearaging definition}). We find the following, 

\begin{equation}
   \begin{split}
       \langle\Dot{J}_1\rangle =&   -\frac{8 m_1^5 m_2^2 \xi ^4 \chi  \left(3 \chi ^2+1\right)}{5 M^6 } \mathcal{J}_{\chi}\\
   \langle\Dot{m}_1\rangle =&  -\frac{8\epsilon  m_1^5 m_2^2 \xi ^5 \chi  \left(3 \chi ^2+1\right)}{5 M^7
   }\mathcal{M}_{\chi}
   \end{split}
\end{equation}
where the enhancement factors $\mathcal{M}_{\chi}$ and $\mathcal{J}_{\chi}$ are expressed as follows,

\begin{equation}
   \begin{split}
       \mathcal{J}_{\chi} =& \frac{\left(3 e_t^4+24 e_t^2+8\right)}{8\left(1-e_t^2\right)^{9/2} }\\
   \mathcal{M}_{\chi} =& \frac{\left(5 e_t^6+90 e_t^4+120 e_t^2+16\right)}{16 \left(1-e_t^2\right)^6 } 
   \end{split}
\end{equation}

Similar to the non-rotating black holes for rotating black holes too in the limit $e_t\rightarrow 0$ the enhancement factors $\mathcal{M}_{\chi} \rightarrow 1$ and $\mathcal{J} \rightarrow 1$. Then in the leading post-Newtonian order, i.e. $\xi =\xi_{\phi}$ this exactly matches with the results derived for circular orbits in Ref. \cite{Alvi:2001mx, Taylor:2008xy, Chatziioannou:2012gq, Chatziioannou:2016kem}. Assuming $m_1\geq m_2$ and $m_2=qm_1$, $\langle\Dot{m}_1\rangle \propto q^2/(1+q)^7$ and $\langle\Dot{J}_1\rangle \propto  q^2/(1+q)^7$. In EMRI limit this becomes $\langle\Dot{m}_1\rangle \propto q^2$ and $\langle\Dot{J}_1\rangle \propto  q^2$. However, the final result in terms of $\xi_{\phi}$ will require further computations which will be addressed later. In the low eccentricity limit for EMRI similar result was computed in \cite{Isoyama:2021jjd}. 

From these expressions, it is evident the effect of eccentricity sips in through two places. The first one is that the enhancement factor is directly an eccentricity-dependent quantity. Secondly, $\xi$ depends on eccentricity and the GW frequency. Therefore through $\xi$ also eccentricity effects sip in. Note, this leading order expression is exact. There has been no assumption about eccentricity and $\xi$. The leading order has been defined as the leading order in $\xi$, however, since $\xi$ itself depends on $\xi_{\phi}$ these expressions can be expressed in a series of initial eccentricity $e_0$ and $\xi_{\phi}$ to arbitrarily high order as per the requirement.

\begin{figure}[ht]
\includegraphics[width=85mm]{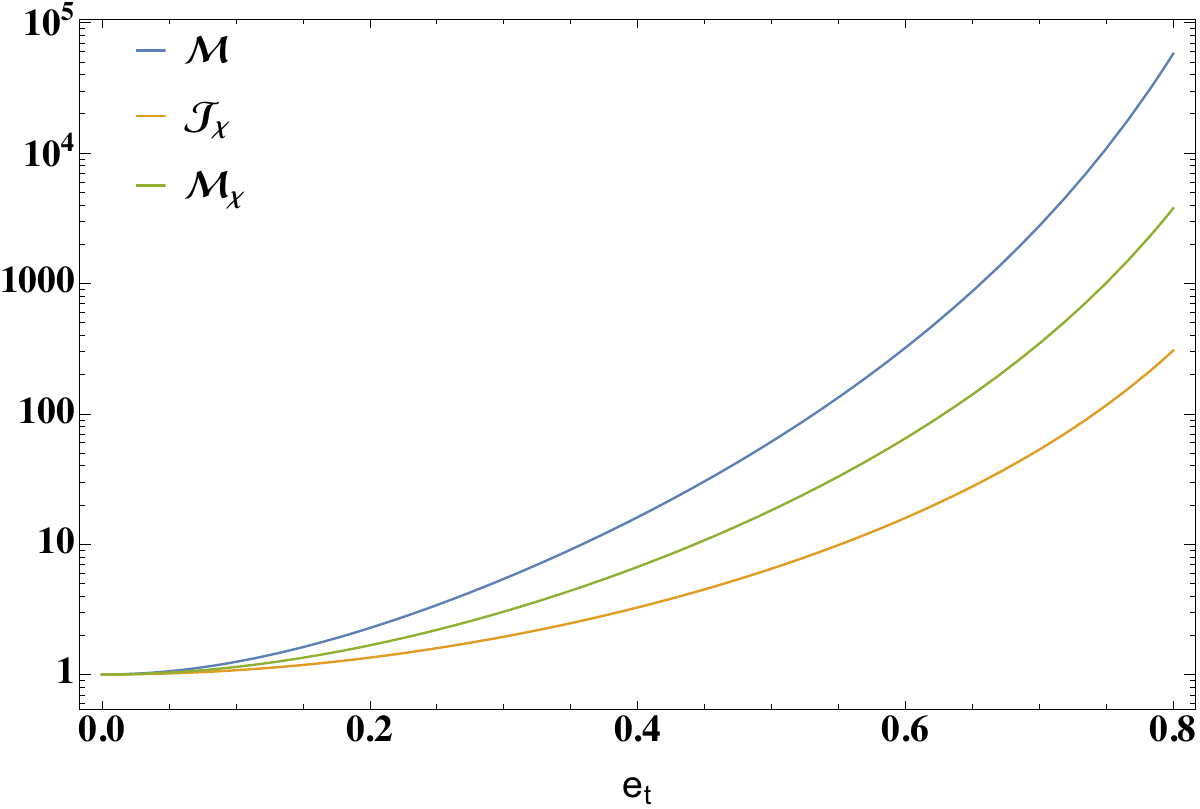}
\caption{\small{In this figure we plot the enhancement factors with respect to the orbital eccentricity. The figure demonstrates that the enhancement is the largest for the rate of change of mass for nonspinning black holes and is the least for the rate of change of angular momentum. However, in all the cases enhancement is significantly large depending on the eccentricity. }}
\label{fig:enhancement} 
\end{figure}

To estimate the impact of the eccentricity on fluxes we plot the enhancement factors in Fig. \ref{fig:enhancement}. In the $y-$axis enhancement factors are plotted with respect to $e_t$ in the $x-$axis. We find that the enhancement is the most significant for non-spinning cases. In the spinning case even though the enhancement is comparatively smaller, it can be $\mathcal{O}(10^3)$ and $\mathcal{O}(10^2)$ for the mass rate and angular momentum rate respectively, for high eccentricity. For circular cases, these functions take the value of the unit. However, for eccentric orbits, these factors can deviate from unit value drastically. Therefore a rate of change of mass and angular momentum can be significantly larger compared to a circular orbit. This plot explicitly demonstrates that in the leading order, the black holes lose angular momentum significantly faster when it is in an eccentric orbit. The same is also true for the mass of the black hole. The only difference is depending on the value of $\epsilon$ the black hole will either increase or decrease its mass. However, in both cases, the magnitude of the rate of change is significantly higher in an eccentric orbit. Note, in the presence of eccentricity both the flux at infinity and horizon gets enhanced. The horizon flux is a $2.5$PN effect. As a result horizon fluxes have a comparatively smaller effect in comparison to leading order flux at infinity. In the eccentric case, the leading order flux at infinity gets enhanced by, $\mathcal{E}_N = (1+73e_t^2/24 + 37e_t^4/96)/(1-e_t^2)^{7/2}$ \cite{Arun:2009mc}. $\mathcal{M}_{\chi}/\mathcal{E}_N$ can take the value $1,\, 1.07033,\, 3.71728,\, 33.909$ respectively for $e_t = .1,\, .5,\, .8,\, .95$. Therefore although the horizon flux is a higher PN effect, it gets a significantly larger enhancement compared to the leading order flux at infinity in the orbital time scale. It implies that the importance of horizon fluxes increases with increasing eccentricity.

\section{eccentricity evolution}
\label{sec:eccentricity evolution}

In this section, we will focus on computing the eccentricity of a binary in terms of binary frequency. In this section, we will construct a prescription that can be used to find the eccentricity in terms of initial eccentricity and frequency. For a non-spinning binary, it is applicable for any PN order and arbitrarily high power of initial eccentricity. For spinning binary, it is applicable to arbitrary order of initial eccentricity and at least up to 1.5PN. 

In Sec. \ref{sec:eccentric intro}, we discussed that in the quasi-Keplerian approach, a parametric solution can be found considering the constants of the motions considering only the conservative part. However, during orbital motion, the system emits GWs that carry away energy and angular momentum. Due to the change in the energy and angular momentum of the orbit, the conserved quantities start to change in the inspiral time scale. This corresponds to the dissipative part of the equation of motion. Conventionally it is tackled by solving the evolution equation of the conserved quantities in the inspiral time scale. For the current work, we only require the evolution equation of the $e_t$ and $n$. Their evolution equation can be expressed as \cite{Arun:2009mc,Klein:2010ti},

\begin{equation}
    \begin{split}
        \left\langle\frac{dn}{dt}\right\rangle =& \frac{\eta x^{11/2}}{m^2(1-e_t^2)^{7/2}}\left(\mathcal{N}_n + x\mathcal{N}_{1} +x^{3/2}\mathcal{N}_{hered}-x^{3/2}\mathcal{N}_{\beta}\right)\\
        \left\langle\frac{de_t}{dt}\right\rangle =& \frac{-e_t \eta x^4}{m(1-e_t^2)^{5/2}}\left(\mathcal{E}_n + x\mathcal{E}_{1} +x^{3/2}\mathcal{E}_{hered}-x^{3/2}\frac{\mathcal{E}_{\beta}}{2e_t^2}\right),
    \end{split}
\end{equation}

where $x = \xi_{\phi}^{2/3}$. Note, that the impact of spin enters through $\mathcal{N}_{\beta}$ and $\mathcal{E}_{\beta}$. Detailed expressions of these terms are provided in the appendix. From now on in the current section, we will suppress the $\langle\rangle$ representing the average, for brevity. The above two equations can be used to find a single equation of $de_t/dx$. The right-hand side of the equation then can be expressed in a series expansion of $e_t$. The coefficients of each term will then depend on $x$. This equation then is solved to find the expression of $e_t$ in terms of an initial eccentricity $e_0$ and $\xi_{\phi}$. In this section we will develop a prescription that can be used to find very high order powers in $e_0$.

To demonstrate the prescription, we will keep up to power $e_t^3$. Then the eccentricity evolution can be expressed as follows,

\begin{eqnarray}
\label{eq:e derivative}
    \frac{de_t}{dx} =& \sum_{n=1,\,{\rm odd}}^{\infty} e_t^n f_n(x) \\
        \frac{de_t}{dx} =& e_t f_1(x) + e_t^3 f_3(x) + \mathcal{O}(e_t^5)
\end{eqnarray}

Note, that we have not explicitly specified the functional expression of $f_n(x)$. Hence, these functions can be expanded in $x$ to arbitrarily high PN order as per the requirement. Once the prescription is described, we will substitute them with the required PN accuracy. 

The resulting solution can be found from Eq. (\ref{eq:e derivative}) by integrating the equation on both sides. During the process, we take $e_t \rightarrow e_0$, i.e. the initial eccentricity, when $x \rightarrow x_0 
 =\xi_{\phi,0}^{2/3}$, i.e. the reference frequency. For the sake of brevity, we will demonstrate the derivation of $\mathcal{O}(e_0^3)$ only. However, we will provide the result up to $\mathcal{O}(e_0^5)$. We find,

\begin{equation}
    \begin{split}
        \int_{e_0}^{e_t}\frac{de_t}{e_t} =& \int_{x_0}^{x}dx \left(f_1(x) + e_t^2 f_3(x) + \mathcal{O}(e_t^3)\right)\\
    \ln(\frac{e_t}{e_0}) =&  \int_{x_0}^{x}dx f_1(x) + \int_{x_0}^{x}dx e_t^2 f_3(x) + \mathcal{O}(e_t^3)\\ 
    e_t =&  e_0e^{\int_{x_0}^{x}dx f_1(x)}  e^{\int_{x_0}^{x}dx e_t^2 f_3(x) + \mathcal{O}(e_t^3)}.
    \end{split}
\end{equation}

Note, that the first integral on the right-hand side is independent of the eccentricity while the second one is dependent. Therefore the first integral can be computed. Therefore the expression can be rearranged as, 

\begin{equation}
    e_t =  e_0\frac{e^{F_1{(x})} }{e^{F_1{(x_0)}} } e^{\int_{x_0}^{x}dx e_t^2 f_3(x) + \mathcal{O}(e_t^3)}.
\end{equation}

The second integral has $e_t$ inside the integral. Therefore, without the exact knowledge of $e_t$ in terms of $x$ this integral can not be computed exactly. However, the eccentricity $e_t$ inside the second integral can be replaced with the above equation and as a result, a leading order term of the integral can be computed. Hence, although an exact integral is not computable, an approximate result can be found which is exact to a particular order. This as a result can be used to find the next order term. This can be continued for arbitrary powers of $e_0$. However, we will only derive upto $e_0^5$. But in principle, this can be continued iteratively by considering higher order terms from Eq. (\ref{eq:e derivative}). Since we have not explicitly assumed the functional form of $f_i(x)$, where $i$ represents the power of $e_t$, the individual coefficient can have higher post-Newtonian order terms included in it. Therefore, it can be used for very high PN orders. After rearranging the expressions they can be expressed as,

\begin{widetext}
\begin{equation}
    \begin{split}
        \frac{e_t e^{F_1{(x_0)}} }{e_0 e^{F_1{(x})}} =& e^{e_0^2\int_{x_0}^{x}dx \frac{e^{2F_1{(x})} }{e^{2F_1{(x_0)}} } \left(e^{2\int_{x_0}^{x}d\Bar{x} e_t^2 f_3(\Bar{x}) + \mathcal{O}(e_t^3)}\right)\,\, f_3(x) + \mathcal{O}(e_t^3)}\\
        \frac{e_t e^{F_1{(x_0)}} }{e_0 e^{F_1{(x})}} =& \left(1+e_0^2\int_{x_0}^{x}dx \frac{e^{2F_1{(x})} }{e^{2F_1{(x_0)}} } \left(e^{2\int_{x_0}^{x}d\Bar{x} e_t^2 f_3(\Bar{x}) + \mathcal{O}(e_t^3)}\right)\,\, f_3(x) + \mathcal{O}(e_0^3)\right)\\
        \frac{e_t e^{F_1{(x_0)}} }{e_0 e^{F_1{(x})}} =& \left(1+e_0^2\int_{x_0}^{x}dx \frac{e^{2F_1{(x})} }{e^{2F_1{(x_0)}} } \left(1 + \mathcal{O}(e_0^2)\right)\,\, f_3(x) + \mathcal{O}(e_0^3)\right)
    \end{split}
\end{equation}
\end{widetext}

This as a result boils down to a series of $e_0$, where the coefficients of the expansions are integrals in $x$. This approach can be used consecutively after deriving individual coefficients of a particular order. Interestingly, to find the schematic structure of the coefficients in integral form of an arbitrary power of $e_0$ it is not required to know $f_i(x)$. The expression can be derived in terms of the integrals of $f_i(x)$s. 

The above results can be expressed in a further simplified and compact form as,

\begin{equation}
        \begin{split}
            e_t =&\frac{e^{F_1{(x})} }{e^{F_1{(x_0)}} } (e_0+e_0^3\int_{x_0}^{x}dx \frac{e^{2F_1{(x})} }{e^{2F_1{(x_0)}} }  f_3(x) + \mathcal{O}(e_0^4))
        \end{split}
\end{equation}

Continuing the similar approach even higher order expression can be found. For brevity, we demonstrate only the final result,

\begin{equation}
\label{eq:final et}
        \begin{split}
            e_t =& e_0\frac{\mathcal{A}_1(x)}{\mathcal{A}_1(0)} + e_0^3\frac{\mathcal{A}_1(x) }{\mathcal{A}_1(0)^3} \left[\mathcal{A}_3(x)- \mathcal{A}_3(0) \right] +e_0^5\frac{\mathcal{A}_1}{\mathcal{A}_1(0)^5}[ \mathcal{A}_5 (x) - \mathcal{A}_5(0) + \frac{1}{2}\mathcal{A}_3(x)^2 -3 \mathcal{A}_3(x)\mathcal{A}_3(0) + \frac{5}{2}\mathcal{A}_3(0)^2 ] + \mathcal{O}(e_0^7)\\
        \end{split}
\end{equation}

where,

\begin{eqnarray}
\label{eq:A A2 B}
    \mathcal{A}_1(x) =& e^{F_1{(x})} \nonumber
    \\
    \mathcal{A}_3(x) - \mathcal{A}_3(x_0) =& \int_{x_0}^{x}d\Bar{x} e^{2F_1{(\Bar{x}})}   f_3(\Bar{x}) \nonumber\\
    \mathcal{A}_5(x) - \mathcal{A}_5(x_0) =& \int_{x_0}^xd\Bar{x} \Big(2\mathcal{A}_1(\Bar{x})^2 f_3(\Bar{x}) \mathcal{A}_3(\Bar{x}) + \mathcal{A}_1(\Bar{x})^4 f_5(\Bar{x})\Big).
\end{eqnarray}

and $\mathcal{A}_i(0) = \mathcal{A}_i(\xi_{\phi,0}) =\mathcal{A}_i(x_0)$. Here we have found the expression for up to $e_0^5$. However, this approach can be continued to find the higher powers of $e_0$. Note, that the above equations do not depend on which PN order is considered. From higher PN calculations only the corrections to $f_i$s are needed. Then Eq. \ref{eq:A A2 B} can be used to find the eccentricity evolution. In \ref{app:Higher order e0} we provide the approximate expression for $\mathcal{O}(e_0^7)$ and $\mathcal{O}(e_0^9)$ for the nonrotating case. For the current purpose we will use $1.5$PN order expressions of $f_1(x)$, $f_3(x)$ and $f_5(x)$. The expressions used for $\Dot{n}$ and $\Dot{e}_t$, where overdot represents time derivative, are shown in \ref{app:rates}.

\begin{eqnarray}
    f_1(x) =& -\frac{19}{12
   x } \left(1+\frac{(2833-5516 \nu ) x }{3192}+\frac{1}{456} x^{3/2}  (-2452 \beta_a-1776 \beta_b+1131
   \pi )\right)\\
    f_3(x) =& \frac{3323 }{576 x }\left(1+\frac{(472943-653228 \nu )
   x }{159504}+\frac{x^{3/2}   (-353345 \beta_a-270423 \beta_b+159321
   \pi )}{39876}\right)\\
    f_5(x) =& -\frac{225431}{13824 x } \left(1+\frac{(232485915-285412372 \nu )
   x}{37872408}+x^{3/2}  \left(\frac{-6313343 \beta_a-4954393 \beta_b}{450862}+\frac{11212300784165054871261 \pi
   }{1761179687500000000000}\right)\right).
\end{eqnarray}

The dependence on the spin enters through $\beta_a$ and $\beta_b$. Then further computation can be done using Eq. (\ref{eq:A A2 B}). In the following, we show the expressions up to $1.5$PN in terms of $\xi_{\phi}$, 
 
\begin{equation}
    \mathcal{A}_1(\xi_{\phi}) = \xi_{\phi}^{-19/18}(1+\frac{(5516 \nu -2833) \xi_{\phi}^{2/3}}{2016}+\frac{1}{432} \xi_{\phi}  (2452 \beta_a+1776 \beta_b-1131 \pi
   ))
\end{equation}

\begin{equation}
    \begin{split}
        \mathcal{A}_3(\xi_{\phi}) = -\frac{3323}{1824 \xi_{\phi}^{19/9}} \left(1+\frac{19 (576485 \nu +64718)
   \xi_{\phi}^{2/3}}{5443074}-\frac{19 \xi_{\phi}  (890535 \pi -2 (893893 \beta_a+517017
   \beta_b))}{7177680}\right)
    \end{split}
\end{equation}

\begin{equation}
    \begin{split}
        \mathcal{A}_5(\xi_{\phi}) =& \frac{19608707 }{3326976 \xi_{\phi}^{38/9}}\Bigg(1  +\frac{95 (21825056708 \nu +2829020183)
   \xi_{\phi}^{2/3}}{513904993056}  \\
   &+\frac{19 \xi_{\phi}  }{58826121}\left(\frac{59}{5220} (2824399609 \beta_a +1699967901 \beta_b)-\frac{2633309468007384992014369 \pi
   }{169921875000000000}\right)\Bigg)
    \end{split}
\end{equation}

\begin{equation}
  \begin{split}
      \mathcal{A}_1(\xi_{\phi})\mathcal{A}_3(\xi_{\phi})=  &-\frac{3323 }{1824 \xi_{\phi}^{19/6}}\left(1+\frac{(137845708 \nu -34236165)
   \xi_{\phi}^{2/3}}{29029728}+\frac{\xi_{\phi}  (12451319 \beta_a+8192481 \beta_b -5951955 \pi )}{1196280}\right)
  \end{split}
\end{equation}

\begin{equation}
    \begin{split}
        \mathcal{A}_1(\xi_{\phi}) \mathcal{A}_5(\xi_{\phi}) =& \frac{19608707}{3326976 \xi_{\phi} ^{95/18}} \Bigg(1   +\frac{(248534396344 \nu -32386582337) \xi_{\phi}
   ^{2/3}}{36707499504} \\
   &+\frac{  -228607670547148483607319033 \pi \xi_{\phi}}{29987534337890625000000000}+\frac{\xi_{\phi}  (47940155111171875000 \beta_a+30938208727353515625 \beta_b)}{2998753433789062500}\Bigg)
    \end{split}
\end{equation}

\begin{equation}
    \begin{split}
        \mathcal{A}_1(\xi_{\phi})\mathcal{A}_3(\xi_{\phi})^2 = \frac{11042329 }{3326976 \xi_{\phi} ^{95/18}}\left(1 +\frac{(588788564
   \nu -83034223) \xi_{\phi} ^{2/3}}{87089184}+ \frac{\xi_{\phi}  (108675848 \beta_a+68801532 \beta_b-52631895 \pi )}{7177680}\right),
    \end{split}
\end{equation}

where $\nu=\eta$. We are keeping both of the symbols as different texts use these two symbols. We will use them interchangeably. These newly found results are consistent with the leading order result found in Ref. \cite{Yunes:2009yz}. These expressions will be used in the next section to compute the horizon fluxes in series of $e_0$ and $\xi_{\phi}$ and dephasing. In an inspiral, as discussed above, the orbital elemnts such as eccentricity and latus rectum are not fixed quantity. With the loss of GWs these quantities evolve. With the inspiral the frequency of the emitted GW also evolves. Therefore, modelling the waveforms in frequency domain requires the evolution equations of the orbital quantities. Finding the fluxes and orbital quantities in terms of $e_t$ and $\xi_{\phi}$, is not sufficient to construct a frequency domain waveform in the manner of \cite{Moore:2016qxz}. However, once the $e_t$ is expressed in terms of an initial eccentricity and GW frequency $\sim \xi_{\phi}$, stationary phase approximation can be used to find the phase and amplitude of the emitted GW in the frequency domain \cite{Tichy:1999pv}. For that purpose, expressing the fluxes in terms of $\xi_{\phi}$ and ``fixed quantities", such as $e_0$, will help in constructing frequency domain waveforms.

\section{Rate of change of mass and angular momentum from L.O. term}
\label{sec:Rate of change}

In section \ref{sec:Leading order TH}, we found the average leading order fluxes for both the spinning and non-spinning BHs. We showed that eccentricity makes the TH much stronger compared to a circular orbit. However, the expressions were computed in terms of $e_t$ which changes over time due to the GW emission. Expressing the results in terms of $\xi$ is also not useful as it is not directly connected to the frequency of the emitted GW, unlike $\xi_{\phi}$.
 Using the results derived in the previous sections here we will re-express them in terms of the initial eccentricity $e_0$ and $\xi_{\phi}$.
 
 The effect of the eccentricity arises only via the enhancement factors and $\xi$. Therefore we only need to do series expansion of $\xi^6 \mathcal{M}$, $\xi^4 \mathcal{J}_{\chi}$ and $ \xi^5 \mathcal{M}_{\chi}$. After using the dependence of $\xi$ and the enhancement factors on $e_0$ and $\xi_{\phi}$ we find,

\begin{equation}
\begin{split}
        \langle\Dot{m}_1\rangle =& \frac{32 m_1^6 m_2^2 \xi_{\phi}^6}{5  M^8}\left(\mathcal{M}_0 + e_0^2 \mathcal{M}_2 + e_0^4 \mathcal{M}_4 + e_0^6 \mathcal{M}_6 \right)\\
   \langle\Dot{m}_1\rangle_{\chi\neq 0} =&  -\frac{8\epsilon  m_1^5 m_2^2 \chi  \left(3 \chi ^2+1\right)}{5 M^7
   } \xi_{\phi}^5\left(\mathcal{M}_{\chi0} + e_0^2 \mathcal{M}_{\chi2} + e_0^4 \mathcal{M}_{\chi4} + e_0^6 \mathcal{M}_{\chi6} \right)     \\
   \langle\Dot{J}_1\rangle =&   -\frac{8 m_1^5 m_2^2  \chi  \left(3 \chi ^2+1\right) \xi_{\phi}^4}{5 M^6 } \left(\mathcal{J}_0 + e_0^2 \mathcal{J}_2 + e_0^4 \mathcal{J}_4 + e_0^6 \mathcal{J}_6 \right)  
\end{split}
\end{equation}

The coefficents $\mathcal{M}_i$, $\mathcal{M}_{\chi i}$, and $\mathcal{J}_i$ can be expressed in postNewtonian expansion. Hence, each of these coefficients is a series expansion in $\xi_{\phi}$. We find,

\begin{equation}
    \begin{split}
    \mathcal{M}_0 =& 1-18 \xi_{\phi} ^{2/3} \\
    \mathcal{M}_{\chi0} =& 1-15 \xi_{\phi} ^{2/3} \\
    \mathcal{J}_0 =& 1-12 \xi_{\phi} ^{2/3}
    \end{split}
\end{equation}

\begin{equation}
    \begin{split}
    \mathcal{M}_2 =& \frac{23}{\mathcal{A}_1(0)^2 \xi_{\phi} ^{19/9}} \left(1+\left(\frac{197 \eta }{36}-\frac{500615}{23184}\right) \xi_{\phi} ^{2/3}-\frac{377 \pi  \xi_{\phi} }{72}\right) \\
    \mathcal{M}_{\chi2} =& \frac{27}{2 \mathcal{A}_1(0)^2 \xi_{\phi} ^{19/9}} \left(1+\left(\frac{197 \eta }{36}-\frac{19073}{1008}\right) \xi_{\phi} ^{2/3} +\left(\frac{613\beta_a}{54}+\frac{74\beta_b}{9}-\frac{377 \pi   }{72} \right) \xi_{\phi}\right)\\
    \mathcal{J}_2 =& \frac{15}{2 \mathcal{A}_1(0)^2 \xi_{\phi} ^{19/9}} \left(1+\left(\frac{197 \eta }{36}-\frac{82709}{5040}\right) \xi_{\phi} ^{2/3}+\left(\frac{613\beta_a}{54}+\frac{74\beta_b}{9}-\frac{377 \pi}{72} \right)\xi_{\phi} \right)
    \end{split}
\end{equation}

\begin{equation}
    \mathcal{M}_4 = \frac{87731 }{912 \mathcal{A}_1(0)^4 \xi_{\phi} ^{38/9}}\left(1+\frac{(16047208268 \nu -34575397879) \xi_{\phi} ^{2/3}}{1149627024}- \frac{  2459723415 \pi
   }{189498960}\xi_{\phi}\right)
\end{equation}

\begin{equation}
     \mathcal{M}_{\chi4} = \frac{13641 }{608 \mathcal{A}_1(0)^4 \xi_{\phi} ^{38/9}}\left(1+\frac{(157727492
   \nu -293973805) \xi_{\phi} ^{2/3}}{8511984} +\frac{\xi_{\phi}  (365520958 \beta_a+279618942 \beta_b-164842035 \pi )}{9821520}\right) 
\end{equation}

\begin{equation}
    \mathcal{J}_4 = -\frac{655 }{608 \mathcal{A}_1(0)^4 \xi_{\phi} ^{38/9}}\left(1 +\frac{(974157013-659362340 \nu ) \xi_{\phi} ^{2/3}}{8583120}+\frac{\xi_{\phi}  (-41087786 \beta_a-34716714 \beta_b+17699745 \pi
   )}{282960}\right)
\end{equation}

\begin{equation}
    \begin{split}
        \mathcal{M}_6 =& -\frac{5862821 }{138624 \mathcal{A}_1(0)^6 \xi_{\phi} ^{19/3}}\Bigg(1 + \frac{(5462576860497-4428488210692
   \nu ) \xi_{\phi} ^{2/3}}{204870417024} + \frac{70720269710659580653919917 \pi }{3984886148437500000000000} \xi_{\phi}  \Bigg)
    \end{split}
\end{equation}

\begin{equation}
    \begin{split}
        \mathcal{M}_{\chi6} =& -\frac{2369655 }{92416 \mathcal{A}_1(0)^6 \xi_{\phi} ^{19/3} } \Bigg(1 + \frac{17
   (9084331932 \nu -15050525335) \xi_{\phi} ^{2/3}}{11829317760} \\
   & +\xi_{\phi}  \left(\frac{185481144929 \beta_a +122831019561 \beta_b}{8246399400}-\frac{5990691756574973175575663 \pi }{536874960937500000000000}\right)\Bigg)
    \end{split}
\end{equation}

\begin{equation}
    \begin{split}
        \mathcal{J}_6 =& \frac{1157705 }{92416 \mathcal{A}_1(0)^6 \xi_{\phi} ^{19/3}}\Bigg(1+ \frac{(73579104553-56767003748 \nu )
   \xi_{\phi} ^{2/3}}{8090968704} \\
   &+\xi_{\phi}  \left(\frac{-11940702143 \beta_a-17960217927 \beta_b }{2417288040}+\frac{282538595290616058525221 \pi }{157375523437500000000000}\right)\Bigg)
    \end{split}
\end{equation}

These expressions of rates of change of mass and angular momentum of the black hole induce an exchange of energy and angular momentum between the orbit and the black hole due to the conservation of total energy and angular momentum. Therefore these are nothing but the fluxes of orbital energy and orbital angular momentum, namely $-\frac{dE}{dt}$ and $-\frac{dL}{dt}$. This rate of change of orbital conserved quantities, therefore, backreacts on the orbit affecting the inspiral rate of binary and as a result the emitted GW. Note, sub-leading terms in $\xi_{\phi}$ are incomplete as only the leading order term of $\xi$ has been considered. In future work, we will compute the subleading terms in $\xi$ which will lead to completion of the subleading terms in $\xi_{\phi}$.

\section{phasing}

In the previous sections, we computed the tidal heating flux of a BH in an eccentric orbit. We also computed the eccentricity evolution including the spin effects. Using such expressions we found the fluxes solely in terms of frequency and initial eccentricity. With these expressions at hand, we can use them to compute the leading order expression of dephasing in the emitted GW due to such an effect. 

The above expressions are used to analytically obtain a parametric solution for the phase $[\langle \phi\rangle (v), \, t(v)]$ by integrating

\begin{equation}
    \frac{d\langle\phi\rangle}{dv} = \frac{d\langle\phi\rangle}{dt}\frac{dt}{dv} = -\frac{v^3 }{M}\frac{dE(v)}{dv}  \frac{1}{F(v)}
\end{equation}

\begin{equation}
    \frac{dt}{dv} = -\frac{dE(v)}{dv} \frac{1}{F(v)} ,
\end{equation}
where $F(v)$ represents flux. From the solution of the above equation, the dephasing is computed as,

\begin{equation}
   \delta \Psi = 2\pi ft(f) - 2\phi[t(f)].
\end{equation}
For further details we refer the reader to Ref. \cite{Moore:2016qxz}. Following the prescription we find the dephasing for both non-spinning and spinning BH.

\begin{figure*}\includegraphics[width=85mm]{dephasing-StlrMBHBNS.pdf}
\includegraphics[width=85mm]{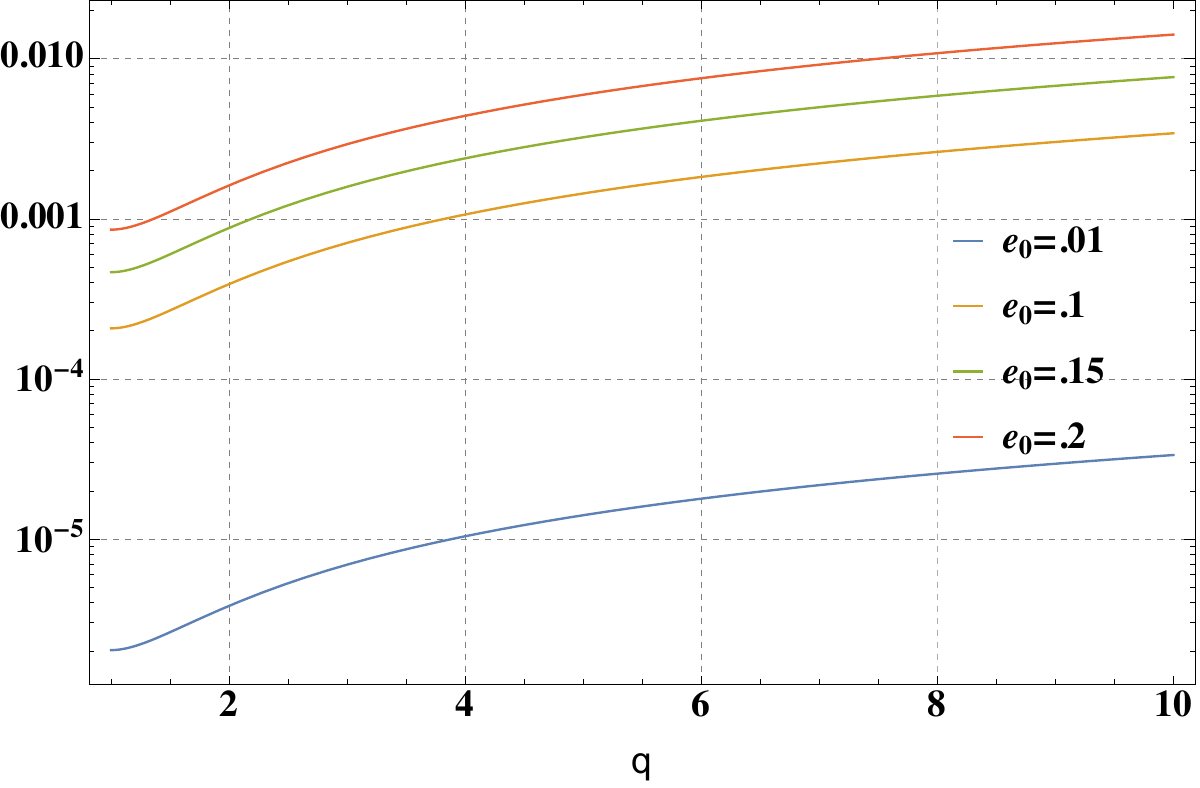}
\caption{In the above figure we show the accumulated dephasing magnitude due to tidal heating of non-spinning black holes in eccentric orbits with respect to the mass ratio $q=m_1/m_2$. In the left panel, we show the dephasing for stellar mass binaries. In the right panel, we show the dephasing for supermassive black hole binaries with $M=2\times 10^6 M_{\odot}$.}
\label{fig:phase BHBNS}
\end{figure*}

\begin{figure*}\includegraphics[width=85mm]{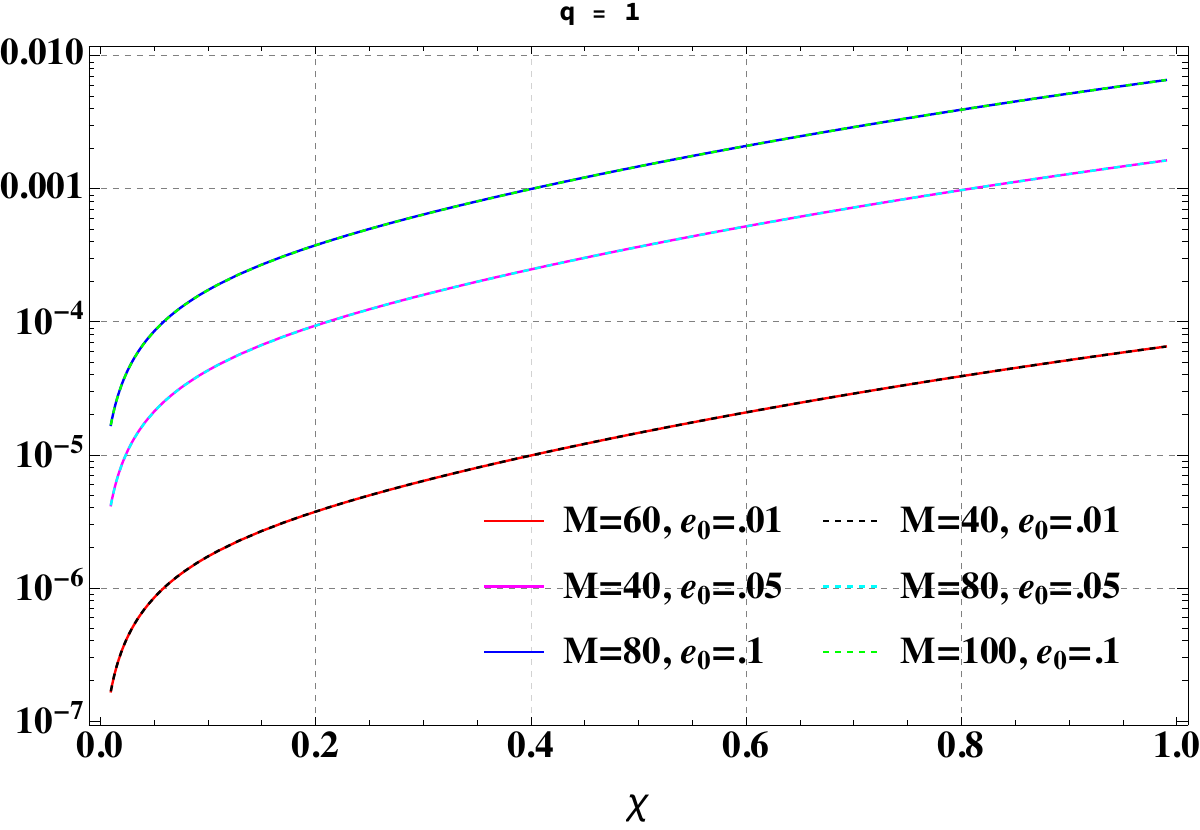}
\includegraphics[width=85mm]{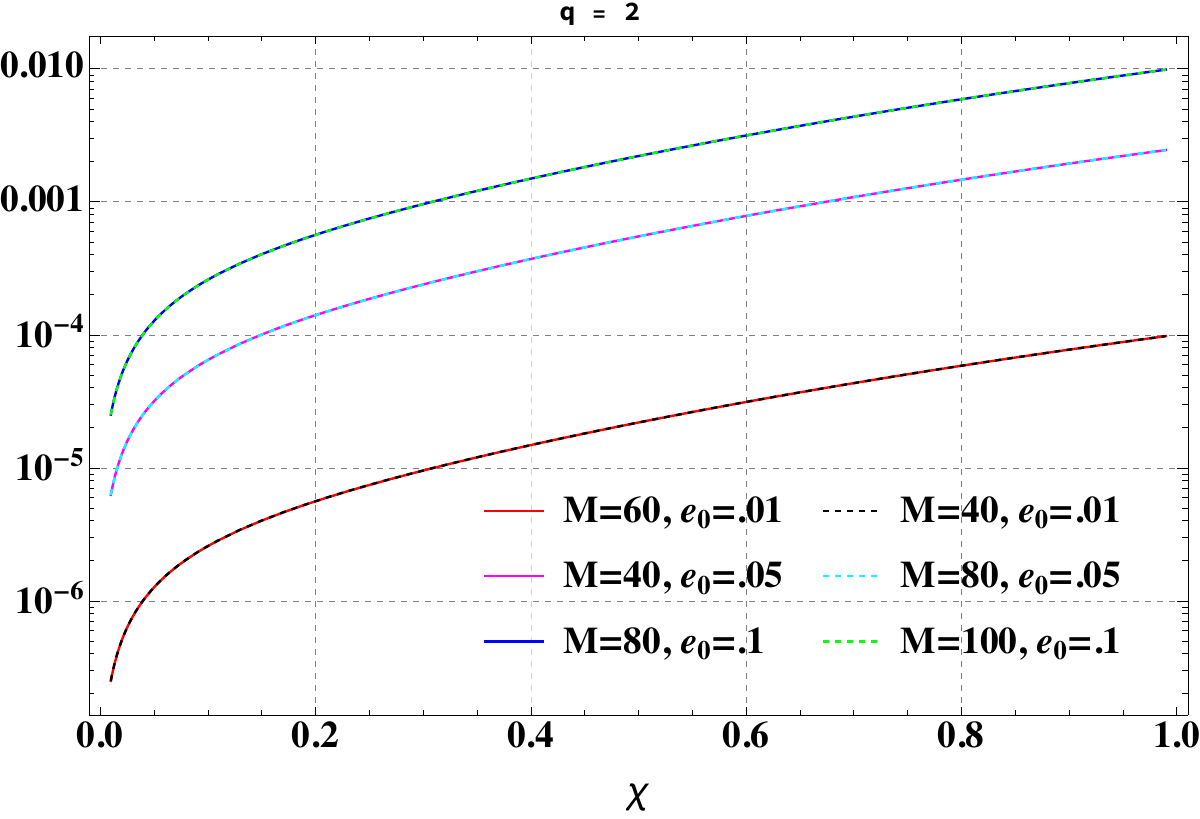}
\includegraphics[width=85mm]{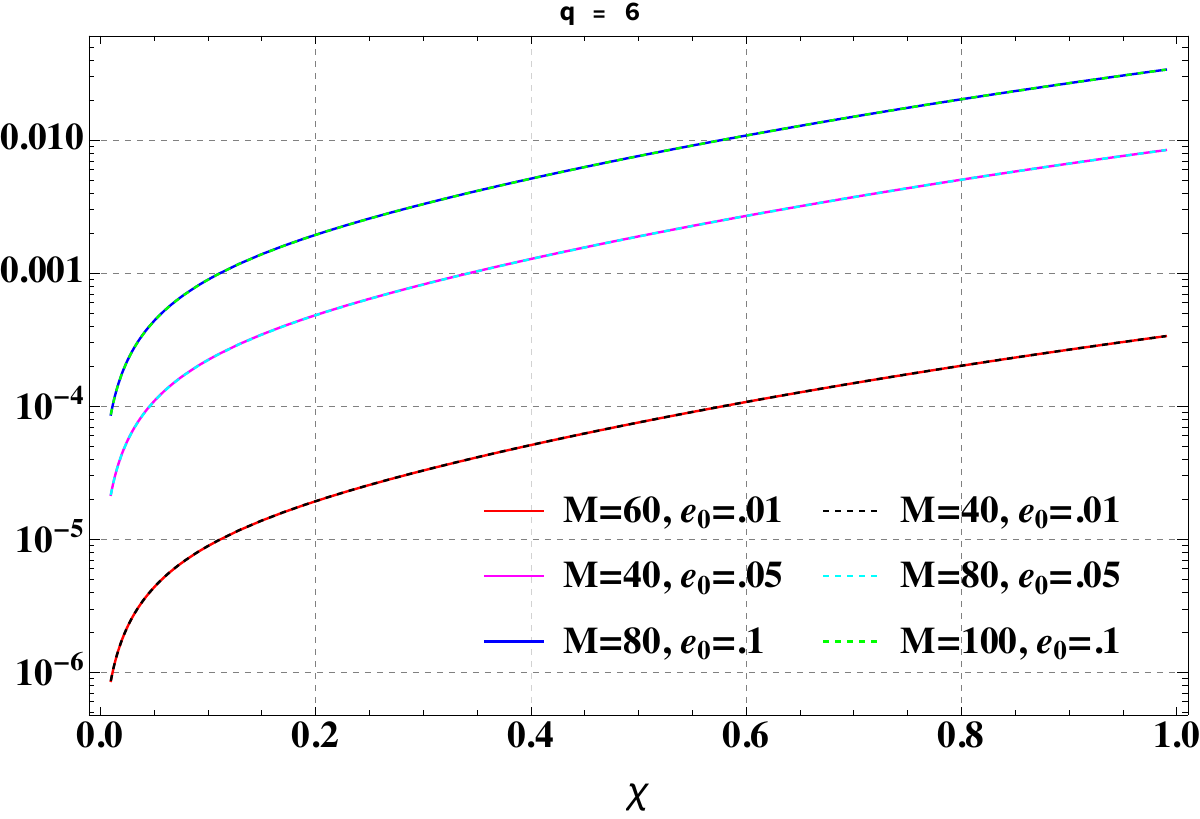}
\includegraphics[width=85mm]{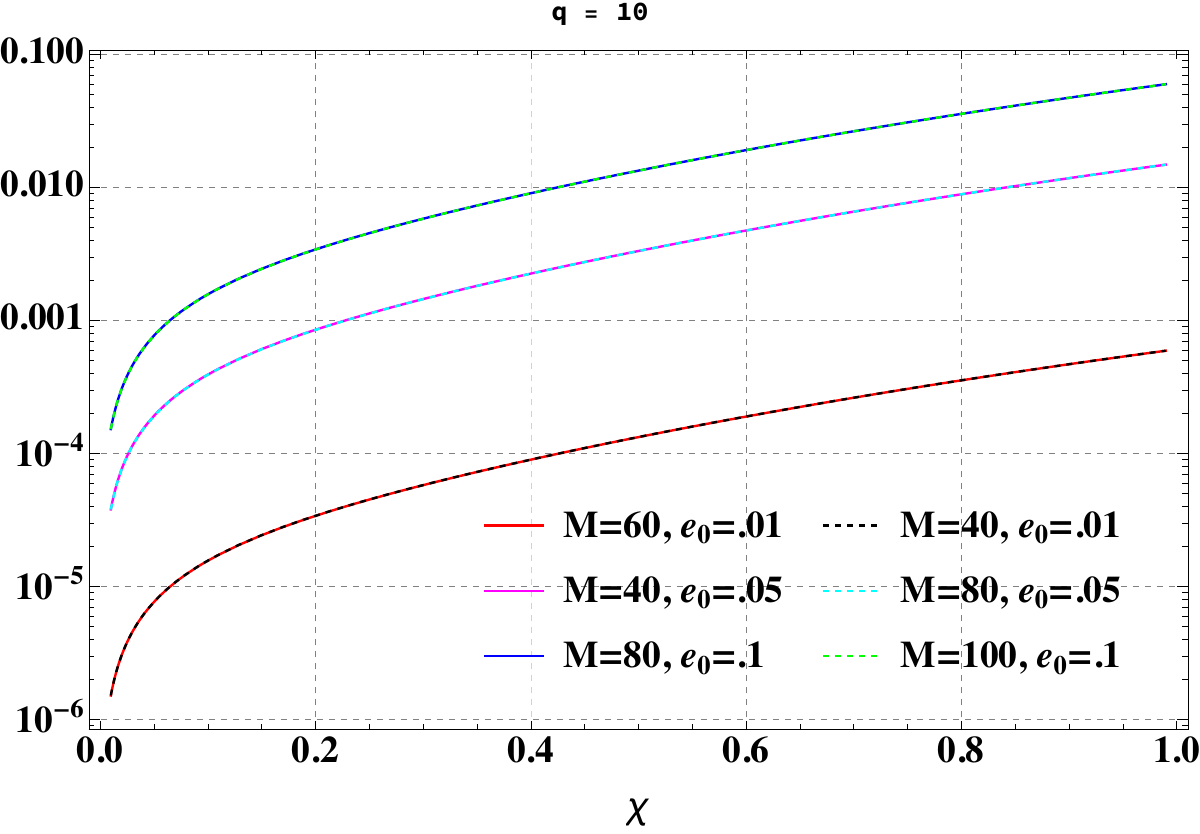}
\caption{In the above figure we demonstrate how the accumulated dephasing magnitude varies with the spin of the BH. We have considered both of the components in the binary to have equal spin $(\chi)$. In a particular figure different curve represents different values of mass and initial eccentricity keeping the mass ratio fixed. In different panels the mass ratio is taken to be different and labeled as $q$.}
\label{fig:phase StlrMBHB spinning}
\end{figure*}

\begin{figure*}
\includegraphics[width=85mm]{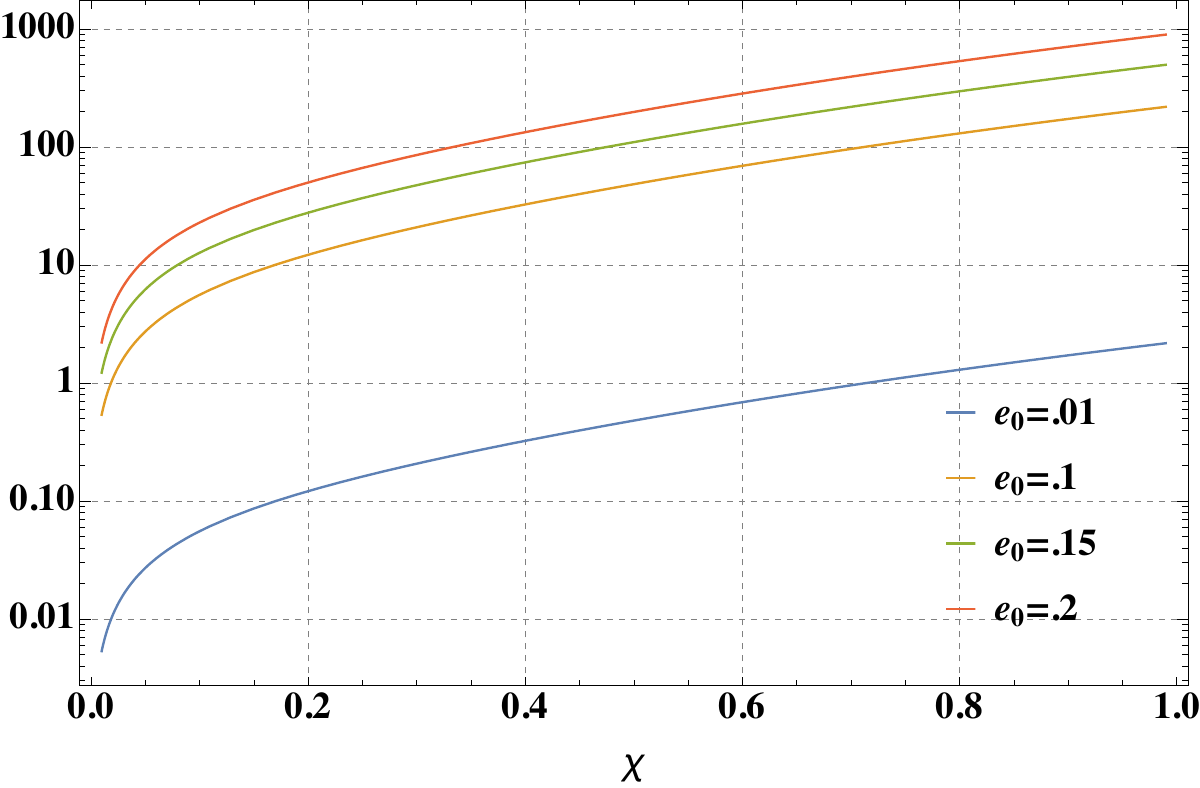}
\includegraphics[width=85mm]{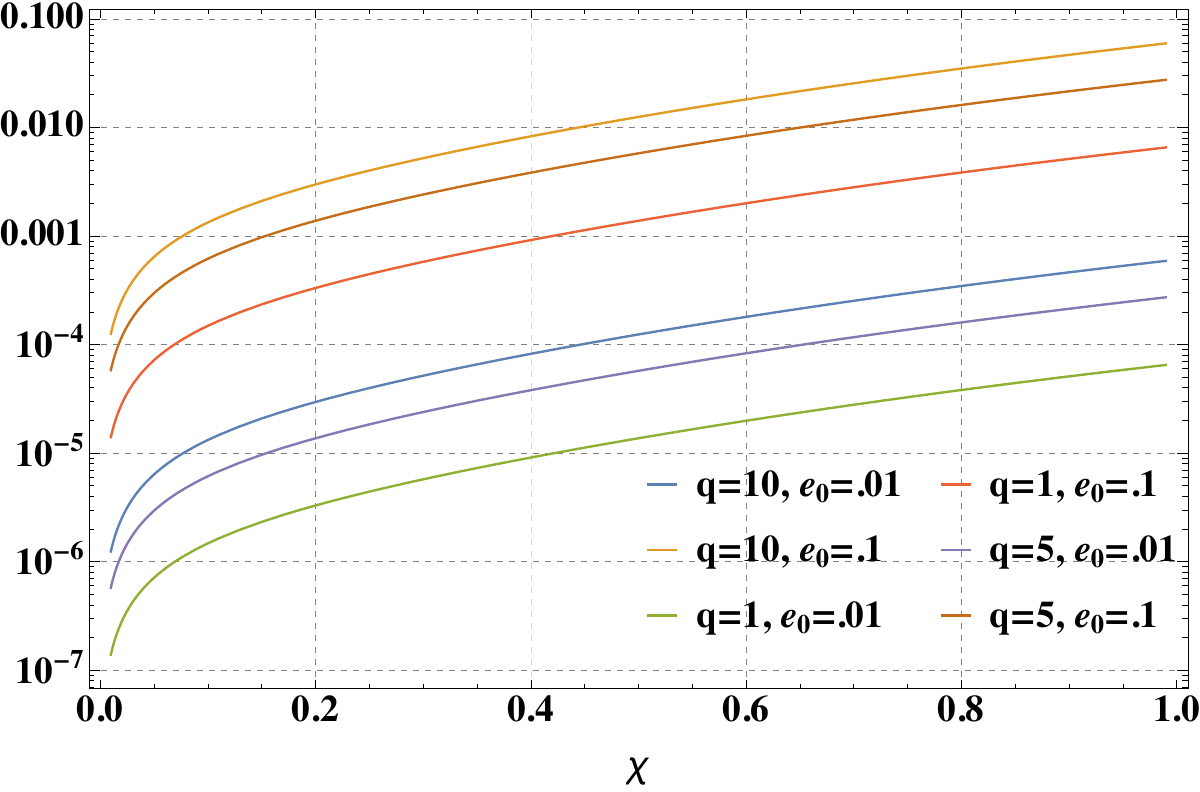}
\caption{In the left panel we show accumulated dephasing magnitude for EMRI with varied initial eccentricity and spin. We considered the supermassive body with $10^6 M_{\odot}$ and the small body with mass $30 M_{\odot}$. In the right panel, supermassive BH binaries have been considered. The total mass is $M=2\times 10^6 M_{\odot}$. Their spins are considered to be equal, and represented by $\chi$.}
\label{fig:phase for spinning EMRI and SMBHB}
\end{figure*}

\begin{equation}
    \delta \Psi_{e_0,\,{\rm TH}, \chi = 0} = - \frac{621 m_1^6 m_2^2 e_0^2 }{608 M^8 \mathcal{A}_1(0)^2 \eta ^3 v^{10/3}} \left(1 +\frac{438655 e_0^2}{608304 \mathcal{A}_1(0)^2
   v^{19/3}}  -\frac{29314105 e_0^4}{229561344 \mathcal{A}_1(0)^4 v^{38/3}}  \right) + 1 \leftrightarrow 2
\end{equation}

\begin{equation}
   \delta \Psi_{e_0,\,{\rm TH}, \chi\neq 0} = \frac{3645 \epsilon m_1^5 m_2^2 \chi_1 \left(3 \chi_1^2+1\right)  e_0^2 }{68096 M^7\mathcal{A}_1(0)^2 \eta ^3 v^{19/3}}\left(1+ \frac{31829 e_0^2}{64296 \mathcal{A}_1(0)^2
   v^{19/3}} -\frac{204785 e_0^4}{762432 \mathcal{A}_1(0)^4 v^{38/3}}\right) + 1 \leftrightarrow 2,
\end{equation}
where $\eta = m_1 m_2/(m_1 + m_2)^2 $ and $1 \leftrightarrow 2$ represents changing the labels of the two blackholes. In Fig. \ref{fig:phase BHBNS} we plot the accumulated dephasing for nonspinning BHs. This represents the total dephasing accumulated in a given frequency band. This is given by:

\begin{equation}\label{eq: accumulated phase}
    \delta \phi = \int^{f_{ISCO}}_{f_0} f \ df \frac{d^2 \delta \Psi(f)}{df^2},
\end{equation}
where $f_{ISCO}$ is the ISCO frequency \cite{Favata:2021vhw}. The dephasing contribution of an effect becomes indistinguishable from the absence of the effect if $\delta\phi^2 \leq \frac{1}{2\rho^2}$, where $\rho$ represents the Signal-to-Noise Ratio (SNR) \cite{Flanagan:1997kp,Lindblom:2008cm}. Consequently, the dephasing parameter $\delta\phi$ remains unobservable unless it surpasses a threshold SNR of approximately $(\sqrt{2}\delta\phi)^{-1}$.
Hence, the dephasing calculated in this manner points toward the required SNR for the effect to be observable.

We show the value of the magnitude of dephasing in different figures. In the left panel of Fig. \ref{fig:phase BHBNS}, non-spinning stellar mass BH binaries are considered whereas in the right panel, non-spinning supermassive BH binaries are considered. For stellar mass binaries $f_0= 4{\rm Hz}$ and SMBH binaries $f_0=1{\rm mHz}$. As can be noticed in both cases the dephasing increases with increasing $e_0$ and $q=m_1/m_2$. In Fig. \ref{fig:phase StlrMBHB spinning} we plot the same but for spinning stellar mass BH binaries. The result seems insensitive to the total mass. This is because the impact of the total mass enters only through the expression of ISCO. However, by the time the binary reaches ISCO frequency the eccentric contribution becomes extremely small. As a consequence only the early phase of the inspiral contributes to the accumulation, resulting in an insensitivity to the mass. Dephasing seems to increase with increasing spin, initial eccentricity, and mass ratio.

In the left panel of Fig. \ref{fig:phase for spinning EMRI and SMBHB} we show the accumulated dephasing for EMRIs. In this case, the dephasing can take a very large value showing the importance of the effect. In the right panel, we show the depashing for spinning SMBH binaries. In this case, the accumulation is respectively smaller. These points that the significant increase in dephasing in EMRI is caused due to the mass ratio of the system. In all other cases, the dephasing is much smaller. However, one should note that next-order corrections may change some of the results.

\section{Discussion and conclusion}

In this paper, we explored the tidal heating of black holes in an eccentric orbit. We only focused on the leading order effect of $\xi$. However, it is of necessary importance to explore the next-order effects. However, currently, there are discrepant results in the literature regarding the next-to-next-leading order results in circular orbits \cite{Saketh:2022xjb}. In Ref. \cite{Saketh:2022xjb} they found that the result does not match with \cite{Chatziioannou:2016kem} in the generic mass ratio case. However in the test mass limit, their result matches with Ref. \cite{Tagoshi:1997jy}. Therefore, this can lead to similar discrepancies in eccentric cases too. This is the reason why we focused only on the leading order result. It is very important to look into the details of the discrepancies and come up with a possible resolution. We will look into this aspect in the future while investigating the next-to-leading order and next-to-next leading order effect in an eccentric orbit. This aspect needs to be studied in detail as the next-generation detectors and space-based detectors will be sensitive to small deviations. Inadequate modeling therefore can lead to a wrong inference of the parameters. Secondly, sources like extreme mass ratio inspirals are expected to have large eccentricities. These sources also demonstrate strong tidal heating effects in the observable band. Therefore, it is crucial that the horizon fluxes are modeled appropriately so that the waveform models reach the required accuracy during observation.

In the process, we also explored the eccentricity evolution of a binary. We developed a prescription that can be used iteratively to find very high-order terms in both the post-Newtonian variable as well as the initial eccentricity $e_0$. We provided the general expression for the coefficient of $\mathcal{O}(e_0^n)$, for up to $n=5$. This expression is valid arbitrary higher PN correction in the functions $f_i(x)$ as long as Eq. (\ref{eq:e derivative}) stays unchanged. With more updated expressions for $f_i(x)$ the correction in the eccentricity evolution can be computed in a straightforward manner using the expression provided here in Eq. (\ref{eq:final et}). In the appendix, we also provide an approximate expression for $\mathcal{O}(e_0^n)$ for arbitrary $n$ and use it to compute the expression for $n=7,9$ in the non-spinning limit. With this result, we also computed dephasing and showed the dephasing values for small eccentricities.

In the current work, we found that the eccentricity effects can change the horizon fluxes significantly from circular orbit results. This is primarily due to the enhancement factor. Depending on the value of the eccentricity, the enhancement factors for horizon energy flux can be as large as $10^4$ for non-spinning bodies and $10^3$ for spinning bodies. We also computed dephasing using the flux expressions. In EMRIs, the effect is very strong. However, in other systems, the dephasing is small implying a large SNR will be required to observe the effect in those systems.

There are certain limitations to the current work. Firstly Eq. (\ref{eq:xi to xiphi}) does not involve spin. Hence, a quasi-Keplerian approach needs to be performed to find corresponding changes and as a consequence in the eccentricity evolution. Secondly, we only computed the rate of change of mass and angular momentum in the leading order of $\xi$. However, using Eq. (\ref{eq:xi to xiphi}) post leading order in $\xi_{\phi}$ result is also computed. Hence, these post-leading order terms in fluxes are incomplete. However, the phase expression here is computed only to the leading order in $\xi_{\phi}$.
Post-leading order results need to be calculated for future purposes.

\section*{Acknowledgement}

We are extremely thankful to K. G. Arun for his comments and suggestions. We thank Eric Poisson for reading an earlier draft. We also thank Chandra Kant Mishra and Samanwaya Mukherjee for useful comments.

\appendix

\section{Expressions for rates}
\label{app:rates}

In this section we will explicitly demonstrate the expression for rate of change of eccentricity. The required expressions are as follows \cite{Arun:2009mc, Klein:2010ti}:

\begin{widetext}
    \begin{eqnarray}
\mathcal{N}_n =&\frac{1}{5}(96 + 292 e_t^2 + 37e_t^4)\\
\mathcal{N}_{1} =&  \frac{1}{(1-e_t^2)}(-\frac{4846}{35} -\frac{264\nu}{5} +e_t^2(\frac{5001}{35} -570\nu) +e_t^4(\frac{2489}{4} -\frac{5061\nu}{10}) + e_t^6 (\frac{11717}{280} - \frac{148\nu}{5})\\
\mathcal{N}_{hered} =&  4\pi\frac{96(1-e_t^2)^{7/2}}{5} \varphi(e_t)\\
\mathcal{N}_{\beta} =&  \frac{1}{10(1-e_t^2)^{3/2}}\beta(3088+15528e_t^2+7026 e_t^4+195e_t^6,2160+11720e_t^2 + 5982 e_t^4 +207 e_t^6) .
\end{eqnarray}

\begin{eqnarray}
\mathcal{E}_n =& (\frac{304}{15} + \frac{121}{15} e_t^2)\\
\mathcal{E}_{1} =&  \frac{1}{(1-e_t^2)}(-\frac{939}{35} -\frac{4084\nu}{45} +e_t^2(\frac{29917}{105} -\frac{7753}{30}\nu) +e_t^4(\frac{13929}{280} -\frac{1664\nu}{45})\\
\mathcal{E}_{hered} =& \frac{32}{5}\frac{985(1-e_t^2)^{5/2}}{48}\pi \varphi_e(e_t)\\
\mathcal{E}_{\beta} =& \frac{e_t^2}{15(1-e_t^2)^{3/2}}\beta(13048+12000e_t^2+789e_t^4,9208 + 10026e_t^2+835e_t^4).
\end{eqnarray}
\end{widetext}

Expression for $\varphi(e_t)$ is taken from \cite{Forseth:2015oua}. Expression of $\varphi_e(e_t)$ is provided in \cite{Arun:2009mc} in terms of different functions, from which up to $\mathcal{O}(e_t^4)$ can be computed. The other terms are computed from higher order terms of $\Tilde{\varphi}(e_t)$ in \cite{Tanay:2016zog}. 

\begin{eqnarray}
\varphi(e_t) =& \frac{1}{(1-e_t^2)^5}(1+\frac{1375}{192}e_t^2 + \frac{3935}{768}e_t^4 + \frac{10007}{36864} e_t^6 + \frac{2321}{884736} e_t^8 +\frac{237857}{353894400} e_t^{10} )\\
\varphi_e(e_t) =& \frac{192}{985}\frac{\sqrt{1-e_t^2}}{e_t^2}\left(\sqrt{1-e_t^2} \varphi(e_t) -\Tilde{\varphi}(e_t)\right)\\
\Tilde{\varphi}(e_t) =& 1 + \frac{209}{32}e_t^2 +\frac{2415}{128} e_t^4 +\frac{317166232530923893657
   }{8000000000000000000} e_t^6 +\frac{280916856316939384413751040219
   }{4000000000000000000000000000} e_t^8 ,
\end{eqnarray}

\begin{eqnarray}
 \beta(a,b) =& a \beta_a + b \beta_b\\
    \beta_a =& {\bf \hat{J}.{\bf \zeta}}\\
    \beta_b =& {\bf \hat{J}.{\bf \xi}}\\
    {\bf \zeta} =& {\bf S_1} + {\bf S_2}\\
    {\bf \xi} =& \frac{m_2}{m_1}{\bf S_1} + \frac{m_1}{m_2}{\bf S_2},
\end{eqnarray}

where ${\bf S}_i$ are the individual spins and ${\bf J}$ is the reduced orbital angular momentum discussed in Ref. \cite{Klein:2010ti}. ${\bf J}$ helps encapsulate the effects of spin precession. It can be checked from Eq. (10) in Ref. \cite{Klein:2010ti}, when the spins of the bodies are either aligned or antialigned with the direction perpendicular to the orbital plane, say $\hat{\bf L}$, ${\bf J}$ is also aligned or antialigned with $\hat{\bf L}$. In such a case therefore, $\hat{\bf J} || \hat{\bf L}$. Using the above expressions it is straightforward to compute $f_i(\xi_{\phi})$s.

\section{Higher order terms of $e_0$}
\label{app:Higher order e0}

From the formalism, it is straightforward to check that the term in the expression of $e_t$ corresponding to $e_0^n$ will consist of a term, 

\begin{equation}
    \mathcal{O}(e_0^n) = \frac{\mathcal{A}_1(x)  }{\mathcal{A}_1(0)}  e_0^n \int_{x_0}^x \frac{\mathcal{A}_1(\Bar{x})^{n-1}  }{\mathcal{A}_1(0)^{n-1}} f_{n-1}(\Bar{x}) d\Bar{x} 
      + {\rm other\,\, terms}.  
\end{equation}

Although it is time-consuming to find the complete expression for $\mathcal{O}(e_0^n)$, it is easier to evaluate the above integral to find an approximate result. For exactness rest of the terms must be found in the manner discussed previously. In this way, we will find the approximate result for several higher-order terms below in the nonspinning limit.

\begin{equation}
    f_7(x) = \frac{15718757 }{331776 x}\left(1 +\frac{(49486716031-57934888172 \nu
   ) x}{5281502352}+ \frac{66835188610235794715208146757739 \pi 
   x^{3/2} }{7675174316406250000000000000000}\right)
\end{equation}

\begin{equation}
    f_9(x) = -\frac{1097423579 }{7962624 x}\left(1 +\frac{(1161963898391-1327655382368 \nu )
   x}{92183580636}+ \frac{370344540064784477905382804708442045113261 \pi 
   x^{3/2} }{33490709808349609375000000000000000000000}\right)
\end{equation}

\begin{equation}
    \mathcal{O}(e_0^7) \sim \frac{\mathcal{A}_1(x)  }{\mathcal{A}_1(0)} e_0^7 \int_{x_0}^x \frac{\mathcal{A}_1(\Bar{x})^6  }{\mathcal{A}_1(0)^6}  f_7(\Bar{x}) d\Bar{x} =\frac{\mathcal{A}_1(x) e_0^7 }{\mathcal{A}_1(0)^7}\left[\mathcal{A}_7(x)-\mathcal{A}_7(0)\right]
\end{equation}

\begin{equation}
    \mathcal{O}(e_0^9) \sim \frac{\mathcal{A}_1(x)  }{\mathcal{A}_1(0)}  e_0^9 \int_{x_0}^x \frac{\mathcal{A}_1(\Bar{x})^8  }{\mathcal{A}_1(0)^8} f_9(\Bar{x}) d\Bar{x} =\frac{\mathcal{A}_1(x) e_0^9 }{\mathcal{A}_1(0)^9}\left[\mathcal{A}_9(x)-\mathcal{A}_9(0)\right]
\end{equation}

\begin{equation}
    \mathcal{A}_7(\xi_{\phi})= -\frac{827303 }{165888 \xi_{\phi}^{19/3}}\left(1 +\frac{5 (410996792 \nu
   +70792535) \xi_{\phi}^{2/3}}{337539624} -\frac{161187023829937147104375559726783 \pi 
   \xi_{\phi} }{19389914062500000000000000000000} \right)
\end{equation}

\begin{equation}
    \begin{split}
        \mathcal{A}_9(\xi_{\phi})= \frac{1097423579}{100859904 \xi_{\phi}^{76/9}} \Bigg(1 &+\frac{19 (1035211157330 \nu +188445347933) \xi_{\phi}^{2/3}}{2419818991695}\\
        &-\frac{56618060827525980254742040394856410285632369 \pi 
   \xi_{\phi} }{5048724503608703613281250000000000000000000}\Bigg)
 \end{split}
 \end{equation}

\begin{equation}
    \mathcal{A}_1(\xi_{\phi})\mathcal{A}_7(\xi_{\phi})=  -\frac{827303 }{165888 \xi_{\phi}^{133/18}}\left(1 +\frac{(250196509556 \nu
   -10110875083) \xi_{\phi}^{2/3}}{28353328416} -\frac{635852688189030191313126679180349 \pi 
   \xi_{\phi} }{58169742187500000000000000000000}\right)
\end{equation}

\begin{equation}
    \begin{split}
        \mathcal{A}_1(\xi_{\phi})\mathcal{A}_9(\xi_{\phi})=  \frac{1097423579}{100859904 \xi_{\phi}^{57/6}} \Bigg(1 &+\frac{(280425659939460 \nu +1919912189173)
   \xi_{\phi}^{2/3}}{25811402578080} \\
   &-\frac{7759544673629801249143429835539601142848041 \pi 
   \xi_{\phi} }{560969389289855957031250000000000000000000}\Bigg)
    \end{split}
\end{equation}

\bibliography{references}

\begin{thebibliography}{59}
\expandafter\ifx\csname natexlab\endcsname\relax\def\natexlab#1{#1}\fi
\expandafter\ifx\csname bibnamefont\endcsname\relax
  \def\bibnamefont#1{#1}\fi
\expandafter\ifx\csname bibfnamefont\endcsname\relax
  \def\bibfnamefont#1{#1}\fi
\expandafter\ifx\csname citenamefont\endcsname\relax
  \def\citenamefont#1{#1}\fi
\expandafter\ifx\csname url\endcsname\relax
  \def\url#1{\texttt{#1}}\fi
\expandafter\ifx\csname urlprefix\endcsname\relax\def\urlprefix{URL }\fi
\providecommand{\bibinfo}[2]{#2}
\providecommand{\eprint}[2][]{\url{#2}}

\bibitem[{\citenamefont{Aasi et~al.}(2015)}]{TheLIGOScientific:2014jea}
\bibinfo{author}{\bibfnamefont{J.}~\bibnamefont{Aasi}} \bibnamefont{et~al.}
  (\bibinfo{collaboration}{LIGO Scientific}), \bibinfo{journal}{Class. Quant.
  Grav.} \textbf{\bibinfo{volume}{32}}, \bibinfo{pages}{074001}
  (\bibinfo{year}{2015}), \eprint{1411.4547}.

\bibitem[{\citenamefont{Acernese et~al.}(2015)}]{TheVirgo:2014hva}
\bibinfo{author}{\bibfnamefont{F.}~\bibnamefont{Acernese}} \bibnamefont{et~al.}
  (\bibinfo{collaboration}{VIRGO}), \bibinfo{journal}{Class. Quant. Grav.}
  \textbf{\bibinfo{volume}{32}}, \bibinfo{pages}{024001}
  (\bibinfo{year}{2015}), \eprint{1408.3978}.

\bibitem[{\citenamefont{Abbott
  et~al.}(2019{\natexlab{a}})}]{LIGOScientific:2018mvr}
\bibinfo{author}{\bibfnamefont{B.~P.} \bibnamefont{Abbott}}
  \bibnamefont{et~al.} (\bibinfo{collaboration}{LIGO Scientific, Virgo}),
  \bibinfo{journal}{Phys. Rev.} \textbf{\bibinfo{volume}{X9}},
  \bibinfo{pages}{031040} (\bibinfo{year}{2019}{\natexlab{a}}),
  \eprint{1811.12907}.

\bibitem[{\citenamefont{Abbott
  et~al.}(2021{\natexlab{a}})}]{LIGOScientific:2020ibl}
\bibinfo{author}{\bibfnamefont{R.}~\bibnamefont{Abbott}} \bibnamefont{et~al.}
  (\bibinfo{collaboration}{LIGO Scientific, Virgo}), \bibinfo{journal}{Phys.
  Rev. X} \textbf{\bibinfo{volume}{11}}, \bibinfo{pages}{021053}
  (\bibinfo{year}{2021}{\natexlab{a}}), \eprint{2010.14527}.

\bibitem[{\citenamefont{Abbott
  et~al.}(2019{\natexlab{b}})}]{LIGOScientific:2019fpa}
\bibinfo{author}{\bibfnamefont{B.~P.} \bibnamefont{Abbott}}
  \bibnamefont{et~al.} (\bibinfo{collaboration}{LIGO Scientific, Virgo}),
  \bibinfo{journal}{Phys. Rev.} \textbf{\bibinfo{volume}{D100}},
  \bibinfo{pages}{104036} (\bibinfo{year}{2019}{\natexlab{b}}),
  \eprint{1903.04467}.

\bibitem[{\citenamefont{Abbott et~al.}(2019{\natexlab{c}})}]{Abbott:2018lct}
\bibinfo{author}{\bibfnamefont{B.~P.} \bibnamefont{Abbott}}
  \bibnamefont{et~al.} (\bibinfo{collaboration}{LIGO Scientific, Virgo}),
  \bibinfo{journal}{Phys. Rev. Lett.} \textbf{\bibinfo{volume}{123}},
  \bibinfo{pages}{011102} (\bibinfo{year}{2019}{\natexlab{c}}),
  \eprint{1811.00364}.

\bibitem[{\citenamefont{Cardoso et~al.}(2017)\citenamefont{Cardoso, Franzin,
  Maselli, Pani, and Raposo}}]{Cardoso:2017cfl}
\bibinfo{author}{\bibfnamefont{V.}~\bibnamefont{Cardoso}},
  \bibinfo{author}{\bibfnamefont{E.}~\bibnamefont{Franzin}},
  \bibinfo{author}{\bibfnamefont{A.}~\bibnamefont{Maselli}},
  \bibinfo{author}{\bibfnamefont{P.}~\bibnamefont{Pani}}, \bibnamefont{and}
  \bibinfo{author}{\bibfnamefont{G.}~\bibnamefont{Raposo}},
  \bibinfo{journal}{Phys. Rev. D} \textbf{\bibinfo{volume}{95}},
  \bibinfo{pages}{084014} (\bibinfo{year}{2017}), \bibinfo{note}{[Addendum:
  Phys.Rev.D 95, 089901 (2017)]}, \eprint{1701.01116}.

\bibitem[{\citenamefont{Abbott et~al.}(2017)}]{gw170817}
\bibinfo{author}{\bibfnamefont{B.~P.} \bibnamefont{Abbott}}
  \bibnamefont{et~al.} (\bibinfo{collaboration}{Virgo, LIGO Scientific}),
  \bibinfo{journal}{Phys. Rev. Lett.} \textbf{\bibinfo{volume}{119}},
  \bibinfo{pages}{161101} (\bibinfo{year}{2017}), \eprint{1710.05832}.

\bibitem[{\citenamefont{Abbott et~al.}(2020)}]{Abbott:2020uma}
\bibinfo{author}{\bibfnamefont{B.}~\bibnamefont{Abbott}} \bibnamefont{et~al.}
  (\bibinfo{collaboration}{LIGO Scientific, Virgo}),
  \bibinfo{journal}{Astrophys.\ J.\ Lett.} \textbf{\bibinfo{volume}{892}},
  \bibinfo{pages}{L3} (\bibinfo{year}{2020}), \eprint{2001.01761}.

\bibitem[{\citenamefont{Abbott
  et~al.}(2021{\natexlab{b}})}]{bhns_LIGOScientific:2021qlt}
\bibinfo{author}{\bibfnamefont{R.}~\bibnamefont{Abbott}} \bibnamefont{et~al.}
  (\bibinfo{collaboration}{LIGO Scientific, KAGRA, VIRGO}),
  \bibinfo{journal}{Astrophys. J. Lett.} \textbf{\bibinfo{volume}{915}},
  \bibinfo{pages}{L5} (\bibinfo{year}{2021}{\natexlab{b}}),
  \eprint{2106.15163}.

\bibitem[{\citenamefont{Maggiore et~al.}(2020)\citenamefont{Maggiore, Van
  Den~Broeck, Bartolo, Belgacem, Bertacca, Bizouard, Branchesi, Clesse, Foffa,
  Garc{\'\i}a-Bellido et~al.}}]{maggiore2020science}
\bibinfo{author}{\bibfnamefont{M.}~\bibnamefont{Maggiore}},
  \bibinfo{author}{\bibfnamefont{C.}~\bibnamefont{Van Den~Broeck}},
  \bibinfo{author}{\bibfnamefont{N.}~\bibnamefont{Bartolo}},
  \bibinfo{author}{\bibfnamefont{E.}~\bibnamefont{Belgacem}},
  \bibinfo{author}{\bibfnamefont{D.}~\bibnamefont{Bertacca}},
  \bibinfo{author}{\bibfnamefont{M.~A.} \bibnamefont{Bizouard}},
  \bibinfo{author}{\bibfnamefont{M.}~\bibnamefont{Branchesi}},
  \bibinfo{author}{\bibfnamefont{S.}~\bibnamefont{Clesse}},
  \bibinfo{author}{\bibfnamefont{S.}~\bibnamefont{Foffa}},
  \bibinfo{author}{\bibfnamefont{J.}~\bibnamefont{Garc{\'\i}a-Bellido}},
  \bibnamefont{et~al.}, \bibinfo{journal}{Journal of Cosmology and
  Astroparticle Physics} \textbf{\bibinfo{volume}{2020}}, \bibinfo{pages}{050}
  (\bibinfo{year}{2020}).

\bibitem[{\citenamefont{Reitze et~al.}(2019)}]{Reitze:2019iox}
\bibinfo{author}{\bibfnamefont{D.}~\bibnamefont{Reitze}} \bibnamefont{et~al.},
  \bibinfo{journal}{Bull. Am. Astron. Soc.} \textbf{\bibinfo{volume}{51}},
  \bibinfo{pages}{035} (\bibinfo{year}{2019}), \eprint{1907.04833}.

\bibitem[{\citenamefont{Amaro-Seoane et~al.}(2017)}]{LISA:2017pwj}
\bibinfo{author}{\bibfnamefont{P.}~\bibnamefont{Amaro-Seoane}}
  \bibnamefont{et~al.} (\bibinfo{collaboration}{LISA}) (\bibinfo{year}{2017}),
  \eprint{1702.00786}.

\bibitem[{\citenamefont{Thorne et~al.}(1986)\citenamefont{Thorne, Price, and
  Macdonald}}]{MembraneParadigm}
\bibinfo{author}{\bibfnamefont{K.~S.} \bibnamefont{Thorne}},
  \bibinfo{author}{\bibfnamefont{R.}~\bibnamefont{Price}}, \bibnamefont{and}
  \bibinfo{author}{\bibfnamefont{D.}~\bibnamefont{Macdonald}},
  \emph{\bibinfo{title}{{Black holes: the membrane paradigm}}}
  (\bibinfo{publisher}{Yale University Press}, \bibinfo{year}{1986}).

\bibitem[{\citenamefont{Damour}(1982)}]{Damour_viscous}
\bibinfo{author}{\bibfnamefont{T.}~\bibnamefont{Damour}}, in
  \emph{\bibinfo{booktitle}{{Pulsating relativistic stars}}}
  (\bibinfo{year}{1982}).

\bibitem[{\citenamefont{Poisson}(2009)}]{Poisson:2009di}
\bibinfo{author}{\bibfnamefont{E.}~\bibnamefont{Poisson}},
  \bibinfo{journal}{Phys. Rev.} \textbf{\bibinfo{volume}{D80}},
  \bibinfo{pages}{064029} (\bibinfo{year}{2009}), \eprint{0907.0874}.

\bibitem[{\citenamefont{Cardoso and Pani}(2013)}]{Cardoso:2012zn}
\bibinfo{author}{\bibfnamefont{V.}~\bibnamefont{Cardoso}} \bibnamefont{and}
  \bibinfo{author}{\bibfnamefont{P.}~\bibnamefont{Pani}},
  \bibinfo{journal}{Class. Quant. Grav.} \textbf{\bibinfo{volume}{30}},
  \bibinfo{pages}{045011} (\bibinfo{year}{2013}), \eprint{1205.3184}.

\bibitem[{\citenamefont{Hartle}(1973)}]{Hartle:1973zz}
\bibinfo{author}{\bibfnamefont{J.~B.} \bibnamefont{Hartle}},
  \bibinfo{journal}{Phys. Rev.} \textbf{\bibinfo{volume}{D8}},
  \bibinfo{pages}{1010} (\bibinfo{year}{1973}).

\bibitem[{\citenamefont{Hughes}(2001)}]{Hughes:2001jr}
\bibinfo{author}{\bibfnamefont{S.~A.} \bibnamefont{Hughes}},
  \bibinfo{journal}{Phys. Rev.} \textbf{\bibinfo{volume}{D64}},
  \bibinfo{pages}{064004} (\bibinfo{year}{2001}), \bibinfo{note}{[Erratum:
  Phys. Rev.D88,no.10,109902(2013)]}, \eprint{gr-qc/0104041}.

\bibitem[{\citenamefont{Poisson and Will}(1953)}]{PoissonWill}
\bibinfo{author}{\bibfnamefont{E.}~\bibnamefont{Poisson}} \bibnamefont{and}
  \bibinfo{author}{\bibfnamefont{C.}~\bibnamefont{Will}},
  \emph{\bibinfo{title}{{Gravity: Newtonian, Post-Newtonian, Relativistic}}}
  (\bibinfo{publisher}{Cambridge University Press},
  \bibinfo{address}{Cambridge, UK}, \bibinfo{year}{1953}).

\bibitem[{\citenamefont{Maselli et~al.}(2018)\citenamefont{Maselli, Pani,
  Cardoso, Abdelsalhin, Gualtieri, and Ferrari}}]{Maselli:2017cmm}
\bibinfo{author}{\bibfnamefont{A.}~\bibnamefont{Maselli}},
  \bibinfo{author}{\bibfnamefont{P.}~\bibnamefont{Pani}},
  \bibinfo{author}{\bibfnamefont{V.}~\bibnamefont{Cardoso}},
  \bibinfo{author}{\bibfnamefont{T.}~\bibnamefont{Abdelsalhin}},
  \bibinfo{author}{\bibfnamefont{L.}~\bibnamefont{Gualtieri}},
  \bibnamefont{and} \bibinfo{author}{\bibfnamefont{V.}~\bibnamefont{Ferrari}},
  \bibinfo{journal}{Phys. Rev. Lett.} \textbf{\bibinfo{volume}{120}},
  \bibinfo{pages}{081101} (\bibinfo{year}{2018}), \eprint{1703.10612}.

\bibitem[{\citenamefont{Datta and Bose}(2019)}]{Datta:2019euh}
\bibinfo{author}{\bibfnamefont{S.}~\bibnamefont{Datta}} \bibnamefont{and}
  \bibinfo{author}{\bibfnamefont{S.}~\bibnamefont{Bose}},
  \bibinfo{journal}{Phys. Rev.} \textbf{\bibinfo{volume}{D99}},
  \bibinfo{pages}{084001} (\bibinfo{year}{2019}), \eprint{1902.01723}.

\bibitem[{\citenamefont{Datta et~al.}(2020)\citenamefont{Datta, Brito, Bose,
  Pani, and Hughes}}]{Datta:2019epe}
\bibinfo{author}{\bibfnamefont{S.}~\bibnamefont{Datta}},
  \bibinfo{author}{\bibfnamefont{R.}~\bibnamefont{Brito}},
  \bibinfo{author}{\bibfnamefont{S.}~\bibnamefont{Bose}},
  \bibinfo{author}{\bibfnamefont{P.}~\bibnamefont{Pani}}, \bibnamefont{and}
  \bibinfo{author}{\bibfnamefont{S.~A.} \bibnamefont{Hughes}},
  \bibinfo{journal}{Phys. Rev.} \textbf{\bibinfo{volume}{D101}},
  \bibinfo{pages}{044004} (\bibinfo{year}{2020}), \eprint{1910.07841}.

\bibitem[{\citenamefont{Datta}(2020)}]{Datta:2020rvo}
\bibinfo{author}{\bibfnamefont{S.}~\bibnamefont{Datta}},
  \bibinfo{journal}{Phys. Rev. D} \textbf{\bibinfo{volume}{102}},
  \bibinfo{pages}{064040} (\bibinfo{year}{2020}), \eprint{2002.04480}.

\bibitem[{\citenamefont{Agullo et~al.}(2021)\citenamefont{Agullo, Cardoso, Rio,
  Maggiore, and Pullin}}]{Agullo:2020hxe}
\bibinfo{author}{\bibfnamefont{I.}~\bibnamefont{Agullo}},
  \bibinfo{author}{\bibfnamefont{V.}~\bibnamefont{Cardoso}},
  \bibinfo{author}{\bibfnamefont{A.~D.} \bibnamefont{Rio}},
  \bibinfo{author}{\bibfnamefont{M.}~\bibnamefont{Maggiore}}, \bibnamefont{and}
  \bibinfo{author}{\bibfnamefont{J.}~\bibnamefont{Pullin}},
  \bibinfo{journal}{Phys. Rev. Lett.} \textbf{\bibinfo{volume}{126}},
  \bibinfo{pages}{041302} (\bibinfo{year}{2021}), \eprint{2007.13761}.

\bibitem[{\citenamefont{Chakraborty et~al.}(2021)\citenamefont{Chakraborty,
  Datta, and Sau}}]{Chakraborty:2021gdf}
\bibinfo{author}{\bibfnamefont{S.}~\bibnamefont{Chakraborty}},
  \bibinfo{author}{\bibfnamefont{S.}~\bibnamefont{Datta}}, \bibnamefont{and}
  \bibinfo{author}{\bibfnamefont{S.}~\bibnamefont{Sau}},
  \bibinfo{journal}{Phys. Rev. D} \textbf{\bibinfo{volume}{104}},
  \bibinfo{pages}{104001} (\bibinfo{year}{2021}), \eprint{2103.12430}.

\bibitem[{\citenamefont{Sherf}(2021)}]{Sherf:2021ppp}
\bibinfo{author}{\bibfnamefont{Y.}~\bibnamefont{Sherf}},
  \bibinfo{journal}{Phys. Rev. D} \textbf{\bibinfo{volume}{103}},
  \bibinfo{pages}{104003} (\bibinfo{year}{2021}), \eprint{2104.03766}.

\bibitem[{\citenamefont{Datta and Phukon}(2021)}]{Datta:2021row}
\bibinfo{author}{\bibfnamefont{S.}~\bibnamefont{Datta}} \bibnamefont{and}
  \bibinfo{author}{\bibfnamefont{K.~S.} \bibnamefont{Phukon}},
  \bibinfo{journal}{Phys. Rev. D} \textbf{\bibinfo{volume}{104}},
  \bibinfo{pages}{124062} (\bibinfo{year}{2021}), \eprint{2105.11140}.

\bibitem[{\citenamefont{Sago and Tanaka}(2021)}]{Sago:2021iku}
\bibinfo{author}{\bibfnamefont{N.}~\bibnamefont{Sago}} \bibnamefont{and}
  \bibinfo{author}{\bibfnamefont{T.}~\bibnamefont{Tanaka}},
  \bibinfo{journal}{Phys. Rev. D} \textbf{\bibinfo{volume}{104}},
  \bibinfo{pages}{064009} (\bibinfo{year}{2021}), \eprint{2106.07123}.

\bibitem[{\citenamefont{Maggio et~al.}(2021)\citenamefont{Maggio, van~de Meent,
  and Pani}}]{Maggio:2021uge}
\bibinfo{author}{\bibfnamefont{E.}~\bibnamefont{Maggio}},
  \bibinfo{author}{\bibfnamefont{M.}~\bibnamefont{van~de Meent}},
  \bibnamefont{and} \bibinfo{author}{\bibfnamefont{P.}~\bibnamefont{Pani}},
  \bibinfo{journal}{Phys. Rev. D} \textbf{\bibinfo{volume}{104}},
  \bibinfo{pages}{104026} (\bibinfo{year}{2021}), \eprint{2106.07195}.

\bibitem[{\citenamefont{Sago and Tanaka}(2022)}]{Sago:2022bbj}
\bibinfo{author}{\bibfnamefont{N.}~\bibnamefont{Sago}} \bibnamefont{and}
  \bibinfo{author}{\bibfnamefont{T.}~\bibnamefont{Tanaka}}
  (\bibinfo{year}{2022}), \eprint{2202.04249}.

\bibitem[{\citenamefont{Pretorius}(2007)}]{Pretorius:2007nq}
\bibinfo{author}{\bibfnamefont{F.}~\bibnamefont{Pretorius}}
  (\bibinfo{year}{2007}), \eprint{0710.1338}.

\bibitem[{\citenamefont{Sasaki and Tagoshi}(2003)}]{Sasaki:2003xr}
\bibinfo{author}{\bibfnamefont{M.}~\bibnamefont{Sasaki}} \bibnamefont{and}
  \bibinfo{author}{\bibfnamefont{H.}~\bibnamefont{Tagoshi}},
  \bibinfo{journal}{Living Rev. Rel.} \textbf{\bibinfo{volume}{6}},
  \bibinfo{pages}{6} (\bibinfo{year}{2003}), \eprint{gr-qc/0306120}.

\bibitem[{\citenamefont{Datta et~al.}(2021)\citenamefont{Datta, Phukon, and
  Bose}}]{Datta:2020gem}
\bibinfo{author}{\bibfnamefont{S.}~\bibnamefont{Datta}},
  \bibinfo{author}{\bibfnamefont{K.~S.} \bibnamefont{Phukon}},
  \bibnamefont{and} \bibinfo{author}{\bibfnamefont{S.}~\bibnamefont{Bose}},
  \bibinfo{journal}{Phys. Rev. D} \textbf{\bibinfo{volume}{104}},
  \bibinfo{pages}{084006} (\bibinfo{year}{2021}), \eprint{2004.05974}.

\bibitem[{\citenamefont{Mukherjee et~al.}(2022)\citenamefont{Mukherjee, Datta,
  Tiwari, Phukon, and Bose}}]{Mukherjee:2022wws}
\bibinfo{author}{\bibfnamefont{S.}~\bibnamefont{Mukherjee}},
  \bibinfo{author}{\bibfnamefont{S.}~\bibnamefont{Datta}},
  \bibinfo{author}{\bibfnamefont{S.}~\bibnamefont{Tiwari}},
  \bibinfo{author}{\bibfnamefont{K.~S.} \bibnamefont{Phukon}},
  \bibnamefont{and} \bibinfo{author}{\bibfnamefont{S.}~\bibnamefont{Bose}},
  \bibinfo{journal}{Phys. Rev. D} \textbf{\bibinfo{volume}{106}},
  \bibinfo{pages}{104032} (\bibinfo{year}{2022}), \eprint{2202.08661}.

\bibitem[{\citenamefont{Taylor and Poisson}(2008)}]{Taylor:2008xy}
\bibinfo{author}{\bibfnamefont{S.}~\bibnamefont{Taylor}} \bibnamefont{and}
  \bibinfo{author}{\bibfnamefont{E.}~\bibnamefont{Poisson}},
  \bibinfo{journal}{Phys. Rev. D} \textbf{\bibinfo{volume}{78}},
  \bibinfo{pages}{084016} (\bibinfo{year}{2008}), \eprint{0806.3052}.

\bibitem[{\citenamefont{Damour and Schaefer}(1988)}]{Damour:1988mr}
\bibinfo{author}{\bibfnamefont{T.}~\bibnamefont{Damour}} \bibnamefont{and}
  \bibinfo{author}{\bibfnamefont{G.}~\bibnamefont{Schaefer}},
  \bibinfo{journal}{Nuovo Cim. B} \textbf{\bibinfo{volume}{101}},
  \bibinfo{pages}{127} (\bibinfo{year}{1988}).

\bibitem[{\citenamefont{Sch{\"a}fer and Wex}(1993)}]{Schafer:1993abc}
\bibinfo{author}{\bibfnamefont{G.}~\bibnamefont{Sch{\"a}fer}} \bibnamefont{and}
  \bibinfo{author}{\bibfnamefont{N.}~\bibnamefont{Wex}}, \bibinfo{journal}{,
  Phys. Lett. A} \textbf{\bibinfo{volume}{174}}, \bibinfo{pages}{196}
  (\bibinfo{year}{1993}).

\bibitem[{\citenamefont{Moore et~al.}(2016)\citenamefont{Moore, Favata, Arun,
  and Mishra}}]{Moore:2016qxz}
\bibinfo{author}{\bibfnamefont{B.}~\bibnamefont{Moore}},
  \bibinfo{author}{\bibfnamefont{M.}~\bibnamefont{Favata}},
  \bibinfo{author}{\bibfnamefont{K.~G.} \bibnamefont{Arun}}, \bibnamefont{and}
  \bibinfo{author}{\bibfnamefont{C.~K.} \bibnamefont{Mishra}},
  \bibinfo{journal}{Phys. Rev. D} \textbf{\bibinfo{volume}{93}},
  \bibinfo{pages}{124061} (\bibinfo{year}{2016}), \eprint{1605.00304}.

\bibitem[{\citenamefont{Damour et~al.}(2004)\citenamefont{Damour, Gopakumar,
  and Iyer}}]{Damour:2004bz}
\bibinfo{author}{\bibfnamefont{T.}~\bibnamefont{Damour}},
  \bibinfo{author}{\bibfnamefont{A.}~\bibnamefont{Gopakumar}},
  \bibnamefont{and} \bibinfo{author}{\bibfnamefont{B.~R.} \bibnamefont{Iyer}},
  \bibinfo{journal}{Phys. Rev. D} \textbf{\bibinfo{volume}{70}},
  \bibinfo{pages}{064028} (\bibinfo{year}{2004}), \eprint{gr-qc/0404128}.

\bibitem[{\citenamefont{Memmesheimer et~al.}(2004)\citenamefont{Memmesheimer,
  Gopakumar, and Schaefer}}]{Memmesheimer:2004cv}
\bibinfo{author}{\bibfnamefont{R.-M.} \bibnamefont{Memmesheimer}},
  \bibinfo{author}{\bibfnamefont{A.}~\bibnamefont{Gopakumar}},
  \bibnamefont{and} \bibinfo{author}{\bibfnamefont{G.}~\bibnamefont{Schaefer}},
  \bibinfo{journal}{Phys. Rev. D} \textbf{\bibinfo{volume}{70}},
  \bibinfo{pages}{104011} (\bibinfo{year}{2004}), \eprint{gr-qc/0407049}.

\bibitem[{\citenamefont{Peters and Mathews}(1963)}]{Peters:1963ux}
\bibinfo{author}{\bibfnamefont{P.~C.} \bibnamefont{Peters}} \bibnamefont{and}
  \bibinfo{author}{\bibfnamefont{J.}~\bibnamefont{Mathews}},
  \bibinfo{journal}{Phys. Rev.} \textbf{\bibinfo{volume}{131}},
  \bibinfo{pages}{435} (\bibinfo{year}{1963}).

\bibitem[{\citenamefont{Alvi}(2001)}]{Alvi:2001mx}
\bibinfo{author}{\bibfnamefont{K.}~\bibnamefont{Alvi}}, \bibinfo{journal}{Phys.
  Rev.} \textbf{\bibinfo{volume}{D64}}, \bibinfo{pages}{104020}
  (\bibinfo{year}{2001}), \eprint{gr-qc/0107080}.

\bibitem[{\citenamefont{Forseth}(2016)}]{Forseth_thesis}
\bibinfo{author}{\bibfnamefont{E.~R.} \bibnamefont{Forseth}}, Ph.D. thesis
  (\bibinfo{year}{2016}),
  \urlprefix\url{https://www.proquest.com/dissertations-theses/high-precision-extreme-mass-ratio-inspirals-black/docview/1805911512/se-2}.

\bibitem[{\citenamefont{Poisson}(2004)}]{Poisson:2004cw}
\bibinfo{author}{\bibfnamefont{E.}~\bibnamefont{Poisson}},
  \bibinfo{journal}{Phys. Rev.} \textbf{\bibinfo{volume}{D70}},
  \bibinfo{pages}{084044} (\bibinfo{year}{2004}), \eprint{gr-qc/0407050}.

\bibitem[{\citenamefont{Saketh et~al.}(2023)\citenamefont{Saketh, Steinhoff,
  Vines, and Buonanno}}]{Saketh:2022xjb}
\bibinfo{author}{\bibfnamefont{M.~V.~S.} \bibnamefont{Saketh}},
  \bibinfo{author}{\bibfnamefont{J.}~\bibnamefont{Steinhoff}},
  \bibinfo{author}{\bibfnamefont{J.}~\bibnamefont{Vines}}, \bibnamefont{and}
  \bibinfo{author}{\bibfnamefont{A.}~\bibnamefont{Buonanno}},
  \bibinfo{journal}{Phys. Rev. D} \textbf{\bibinfo{volume}{107}},
  \bibinfo{pages}{084006} (\bibinfo{year}{2023}), \eprint{2212.13095}.

\bibitem[{\citenamefont{Chatziioannou et~al.}(2013)\citenamefont{Chatziioannou,
  Poisson, and Yunes}}]{Chatziioannou:2012gq}
\bibinfo{author}{\bibfnamefont{K.}~\bibnamefont{Chatziioannou}},
  \bibinfo{author}{\bibfnamefont{E.}~\bibnamefont{Poisson}}, \bibnamefont{and}
  \bibinfo{author}{\bibfnamefont{N.}~\bibnamefont{Yunes}},
  \bibinfo{journal}{Phys. Rev.} \textbf{\bibinfo{volume}{D87}},
  \bibinfo{pages}{044022} (\bibinfo{year}{2013}), \eprint{1211.1686}.

\bibitem[{\citenamefont{Chatziioannou et~al.}(2016)\citenamefont{Chatziioannou,
  Poisson, and Yunes}}]{Chatziioannou:2016kem}
\bibinfo{author}{\bibfnamefont{K.}~\bibnamefont{Chatziioannou}},
  \bibinfo{author}{\bibfnamefont{E.}~\bibnamefont{Poisson}}, \bibnamefont{and}
  \bibinfo{author}{\bibfnamefont{N.}~\bibnamefont{Yunes}},
  \bibinfo{journal}{Phys. Rev.} \textbf{\bibinfo{volume}{D94}},
  \bibinfo{pages}{084043} (\bibinfo{year}{2016}), \eprint{1608.02899}.

\bibitem[{\citenamefont{Isoyama et~al.}(2022)\citenamefont{Isoyama, Fujita,
  Chua, Nakano, Pound, and Sago}}]{Isoyama:2021jjd}
\bibinfo{author}{\bibfnamefont{S.}~\bibnamefont{Isoyama}},
  \bibinfo{author}{\bibfnamefont{R.}~\bibnamefont{Fujita}},
  \bibinfo{author}{\bibfnamefont{A.~J.~K.} \bibnamefont{Chua}},
  \bibinfo{author}{\bibfnamefont{H.}~\bibnamefont{Nakano}},
  \bibinfo{author}{\bibfnamefont{A.}~\bibnamefont{Pound}}, \bibnamefont{and}
  \bibinfo{author}{\bibfnamefont{N.}~\bibnamefont{Sago}},
  \bibinfo{journal}{Phys. Rev. Lett.} \textbf{\bibinfo{volume}{128}},
  \bibinfo{pages}{231101} (\bibinfo{year}{2022}), \eprint{2111.05288}.

\bibitem[{\citenamefont{Arun et~al.}(2009)\citenamefont{Arun, Blanchet, Iyer,
  and Sinha}}]{Arun:2009mc}
\bibinfo{author}{\bibfnamefont{K.~G.} \bibnamefont{Arun}},
  \bibinfo{author}{\bibfnamefont{L.}~\bibnamefont{Blanchet}},
  \bibinfo{author}{\bibfnamefont{B.~R.} \bibnamefont{Iyer}}, \bibnamefont{and}
  \bibinfo{author}{\bibfnamefont{S.}~\bibnamefont{Sinha}},
  \bibinfo{journal}{Phys. Rev. D} \textbf{\bibinfo{volume}{80}},
  \bibinfo{pages}{124018} (\bibinfo{year}{2009}), \eprint{0908.3854}.

\bibitem[{\citenamefont{Klein and Jetzer}(2010)}]{Klein:2010ti}
\bibinfo{author}{\bibfnamefont{A.}~\bibnamefont{Klein}} \bibnamefont{and}
  \bibinfo{author}{\bibfnamefont{P.}~\bibnamefont{Jetzer}},
  \bibinfo{journal}{Phys. Rev. D} \textbf{\bibinfo{volume}{81}},
  \bibinfo{pages}{124001} (\bibinfo{year}{2010}), \eprint{1005.2046}.

\bibitem[{\citenamefont{Yunes et~al.}(2009)\citenamefont{Yunes, Arun, Berti,
  and Will}}]{Yunes:2009yz}
\bibinfo{author}{\bibfnamefont{N.}~\bibnamefont{Yunes}},
  \bibinfo{author}{\bibfnamefont{K.~G.} \bibnamefont{Arun}},
  \bibinfo{author}{\bibfnamefont{E.}~\bibnamefont{Berti}}, \bibnamefont{and}
  \bibinfo{author}{\bibfnamefont{C.~M.} \bibnamefont{Will}},
  \bibinfo{journal}{Phys. Rev. D} \textbf{\bibinfo{volume}{80}},
  \bibinfo{pages}{084001} (\bibinfo{year}{2009}), \bibinfo{note}{[Erratum:
  Phys.Rev.D 89, 109901 (2014)]}, \eprint{0906.0313}.

\bibitem[{\citenamefont{Tichy et~al.}(2000)\citenamefont{Tichy, Flanagan, and
  Poisson}}]{Tichy:1999pv}
\bibinfo{author}{\bibfnamefont{W.}~\bibnamefont{Tichy}},
  \bibinfo{author}{\bibfnamefont{E.~E.} \bibnamefont{Flanagan}},
  \bibnamefont{and} \bibinfo{author}{\bibfnamefont{E.}~\bibnamefont{Poisson}},
  \bibinfo{journal}{Phys. Rev.} \textbf{\bibinfo{volume}{D61}},
  \bibinfo{pages}{104015} (\bibinfo{year}{2000}), \eprint{gr-qc/9912075}.

\bibitem[{\citenamefont{Favata et~al.}(2022)\citenamefont{Favata, Kim, Arun,
  Kim, and Lee}}]{Favata:2021vhw}
\bibinfo{author}{\bibfnamefont{M.}~\bibnamefont{Favata}},
  \bibinfo{author}{\bibfnamefont{C.}~\bibnamefont{Kim}},
  \bibinfo{author}{\bibfnamefont{K.~G.} \bibnamefont{Arun}},
  \bibinfo{author}{\bibfnamefont{J.}~\bibnamefont{Kim}}, \bibnamefont{and}
  \bibinfo{author}{\bibfnamefont{H.~W.} \bibnamefont{Lee}},
  \bibinfo{journal}{Phys. Rev. D} \textbf{\bibinfo{volume}{105}},
  \bibinfo{pages}{023003} (\bibinfo{year}{2022}), \eprint{2108.05861}.

\bibitem[{\citenamefont{Flanagan and Hughes}(1998)}]{Flanagan:1997kp}
\bibinfo{author}{\bibfnamefont{E.~E.} \bibnamefont{Flanagan}} \bibnamefont{and}
  \bibinfo{author}{\bibfnamefont{S.~A.} \bibnamefont{Hughes}},
  \bibinfo{journal}{Phys. Rev. D} \textbf{\bibinfo{volume}{57}},
  \bibinfo{pages}{4566} (\bibinfo{year}{1998}), \eprint{gr-qc/9710129}.

\bibitem[{\citenamefont{Lindblom et~al.}(2008)\citenamefont{Lindblom, Owen, and
  Brown}}]{Lindblom:2008cm}
\bibinfo{author}{\bibfnamefont{L.}~\bibnamefont{Lindblom}},
  \bibinfo{author}{\bibfnamefont{B.~J.} \bibnamefont{Owen}}, \bibnamefont{and}
  \bibinfo{author}{\bibfnamefont{D.~A.} \bibnamefont{Brown}},
  \bibinfo{journal}{Phys. Rev. D} \textbf{\bibinfo{volume}{78}},
  \bibinfo{pages}{124020} (\bibinfo{year}{2008}), \eprint{0809.3844}.

\bibitem[{\citenamefont{Tagoshi et~al.}(1997)\citenamefont{Tagoshi, Mano, and
  Takasugi}}]{Tagoshi:1997jy}
\bibinfo{author}{\bibfnamefont{H.}~\bibnamefont{Tagoshi}},
  \bibinfo{author}{\bibfnamefont{S.}~\bibnamefont{Mano}}, \bibnamefont{and}
  \bibinfo{author}{\bibfnamefont{E.}~\bibnamefont{Takasugi}},
  \bibinfo{journal}{Prog. Theor. Phys.} \textbf{\bibinfo{volume}{98}},
  \bibinfo{pages}{829} (\bibinfo{year}{1997}), \eprint{gr-qc/9711072}.

\bibitem[{\citenamefont{Forseth et~al.}(2016)\citenamefont{Forseth, Evans, and
  Hopper}}]{Forseth:2015oua}
\bibinfo{author}{\bibfnamefont{E.}~\bibnamefont{Forseth}},
  \bibinfo{author}{\bibfnamefont{C.~R.} \bibnamefont{Evans}}, \bibnamefont{and}
  \bibinfo{author}{\bibfnamefont{S.}~\bibnamefont{Hopper}},
  \bibinfo{journal}{Phys. Rev. D} \textbf{\bibinfo{volume}{93}},
  \bibinfo{pages}{064058} (\bibinfo{year}{2016}), \eprint{1512.03051}.

\bibitem[{\citenamefont{Tanay et~al.}(2016)\citenamefont{Tanay, Haney, and
  Gopakumar}}]{Tanay:2016zog}
\bibinfo{author}{\bibfnamefont{S.}~\bibnamefont{Tanay}},
  \bibinfo{author}{\bibfnamefont{M.}~\bibnamefont{Haney}}, \bibnamefont{and}
  \bibinfo{author}{\bibfnamefont{A.}~\bibnamefont{Gopakumar}},
  \bibinfo{journal}{Phys. Rev. D} \textbf{\bibinfo{volume}{93}},
  \bibinfo{pages}{064031} (\bibinfo{year}{2016}), \eprint{1602.03081}.

\end{thebibliography}

\end{document}